\documentclass[11pt,a4paper]{article}          
\usepackage{jheppub}

\usepackage{graphicx}
\usepackage{caption}
\usepackage{subcaption}
\usepackage{float}

\usepackage{mathtools}
\usepackage{mathrsfs}
\usepackage{amssymb}
\usepackage{amsmath}
\usepackage{amstext}
\usepackage{physics}

\newcommand{\be}{\begin{equation}}
	\newcommand{\ee}{\end{equation}}

\def\pa{\partial}

\def\lag{\mathscr{L}}

\def\al{\alpha}
\def\sig{\sigma}
\def\del{\delta}

\def\lam{\lambda}
\def\Lam{\Lambda}
\def\gam{\gamma}
\def\Gam{\Gamma}
\def\na{\nabla}
\def\ka{\kappa}
\def\varep{\varepsilon}
\def\ep{\epsilon}

\def\om{\omega}

\def\sgam{\sqrt{- \gam}}

\def\nn{m}

\usepackage{braket}

\title{Viscoelastic hydrodynamics of charged black holes}

\author{Richard A.~Davison and Andr\'{e} Oliveira Pinheiro} 

\affiliation{Department of Mathematics and Maxwell Institute for Mathematical Sciences, Heriot-Watt University,
Edinburgh EH14 4AS, U.K.}

\emailAdd{r.davison@hw.ac.uk}
\emailAdd{ao2012@hw.ac.uk}
 
\abstract{We study the dynamics of an isotropic, planar AdS black hole charged under a pair of two-form gauge potentials. We prove that long wavelength, small amplitude perturbations of this state are governed by the relativistic theory of viscoelastic hydrodynamics. We use this effective theory to identify instabilities in certain regions of parameter space, and show that the dominant instability is always driven by fluctuations of an incoherent density of the higher-form charge. The spacetime we study also arises as a solution to gravity coupled to a pair of massless scalar fields, where part of its dynamics is governed by a simpler effective theory of heat diffusion. We derive this simpler description directly from viscoelastic hydrodynamics, demonstrating that the heat diffusion mode is the collective excitation of the viscoelastic fluid under no-flux boundary conditions.}

\begin{document} 	
\maketitle

\section{Introduction}

\paragraph{}Due to the AdS/CFT correspondence, the dynamics of certain asymptotically AdS black holes are equivalent to those of certain thermal quantum field theories (QFTs) \cite{Maldacena:1997re,Witten:1998zw}. Although in general these dynamics are complicated, there is a significant simplification over large distance and time scales. This is most obvious in the QFT description, where an effective hydrodynamic theory emerges that describes a locally thermal state (see \cite{Kovtun:2012rj} for a pedagogical introduction). Of course, the same simplification should occur in the gravitational description. In the limit where gravity is classical, hydrodynamic fluctuations are suppressed and the effective theories reduce to more familiar theories of fluid dynamics \cite{Kovtun:2003vj}. Furthermore, in this limit the state is expected to thermalise rapidly so that the regime where the effective theory applies is large \cite{Kovtun:2004de}.

\paragraph{}These expectations have been explicitly verified for some classes of planar AdS black holes, where a fluid/gravity correspondence has been demonstrated: the long wavelength dynamics of the black hole metric are governed by a (lower-dimensional) theory of hydrodynamics \cite{Bhattacharyya:2007vjd}. This is a significant simplification, making manifest a hidden, intuitive structure underlying the dynamics of black holes. For example, it implies that the black hole quasinormal modes -- the characteristic small amplitude excitations that control the relaxation back to equilibrium -- at long wavelengths are nothing but the fundamental excitations of a fluid \cite{Policastro:2002se,Policastro:2002tn}. It has also had important consequences and spurred further developments in the study of hydrodynamics itself. In particular, features of hydrodynamic theories that were previously overlooked have been identified thanks to the relative ease with which one can derive these theories explicitly from gravity (for example, see \cite{Son:2009tf,Erdmenger:2008rm,Banerjee:2008th,Torabian:2009qk} for anomalies, \cite{Delacretaz:2021qqu,Armas:2021vku,Amoretti:2018tzw} for damping of pseudo-Goldstones, \cite{Donos:2025jxb,Donos:2025igh} for nearly-critical superfluids and \cite{Gouteraux:2023uff,Davison:2015bea,Blake:2015epa,Blake:2015hxa} for generalised Drude transport in metals).

\paragraph{}The relevant theory of hydrodynamics depends on the symmetries of the state, both spacetime and internal. In the gravitational description, internal symmetries are realised by gauge fields and so there are a rich variety of fluid descriptions of different charged black holes. The fluid/gravity duality is well-established in the simplest case of a black hole charged under a Maxwell potential $A_\mu$, described by the theory of charged, relativistic hydrodynamics for a locally conserved energy-momentum tensor $T^{\mu\nu}$ and U(1) current $J^\mu$ \cite{Erdmenger:2008rm,Banerjee:2008th} (see \cite{Davison:2022vqh,Donos:2022www} for extensions to non-conformal cases). In this work we are instead interested in black holes charged under higher-form potentials, whose fluid limits should be described by hydrodynamic theories with locally conserved higher-form currents. Although seemingly exotic, the importance of such hydrodynamic theories for the description of magnetised plasmas (one-form symmetry in 3 spatial dimensions) \cite{Grozdanov:2016tdf,Hernandez:2017mch,Armas:2018ibg,Armas:2018atq,Armas:2018zbe},  superfluids (anomalous U(1)$\times$U(1)$^{(d-1)}$ symmetry in $d$ spatial dimensions) \cite{Delacretaz:2019brr}, and viscoelastic crystals ($d$ copies of a $(d-1)$-form symmetry) \cite{Grozdanov:2018ewh,Armas:2019sbe}, amongst others, has been recently emphasised. See \cite{Gaiotto:2014kfa,Bhardwaj:2023kri,Iqbal:2024pee} for introductions to higher-form symmetries.

\paragraph{}We study arguably the simplest example of an AdS black brane with higher-form charge: the (3+1)-dimensional isotropic AdS black hole charged under a pair of two-form gauge potentials. According to the discussion above, its long wavelength limit should be described by relativistic viscoelastic hydrodynamics: the effective theory of an energy-momentum tensor $T^{\mu\nu}$ and a pair of two-form currents $J_I^{\mu\nu}$. This theory is named after its main application: a systematic, relativistic extension of the theory of elasticity in crystals, in which the conserved currents arise as a consequence of spontaneous translational symmetry breaking. This hydrodynamic theory has been developed in recent years \cite{Delacretaz:2017zxd,Grozdanov:2018ewh} due to its potential relevance in strongly correlated materials \cite{Delacretaz:2016ivq,Baggioli:2022pyb}. This culminated in \cite{Armas:2019sbe} where this theory was presented up to first order in the hydrodynamic derivative expansion.

\paragraph{}The development of this hydrodynamic theory was accompanied, and informed, by studies of a variety of gravitational models where viscoelastic hydrodynamics was expected to arise: massive gravity-like models \cite{Alberte:2017oqx,Andrade:2019zey,Ammon:2019wci,Ammon:2019apj,Baggioli:2019aqf,Ammon:2020xyv,Baggioli:2020ljz,Li:2024rzd}, Q-lattice models \cite{Amoretti:2016bxs,Amoretti:2017frz,Amoretti:2017axe,Amoretti:2019cef}, inhomogeneous models \cite{Ling:2014saa,Donos:2018kkm,Gouteraux:2018wfe} and numerous others \cite{Jokela:2016xuy,Esposito:2017qpj,Andrade:2017cnc,Li:2018vrz,Wang:2021jfu,Zhong:2022mok,Xia:2024gba}. These models often involved extra complexities due to realising the viscoelastic phase through spontaneous translational symmetry breaking in a more complete microscopic model involving $0$-form U(1) potentials and/or additional scalar fields. Due to this, the comparisons of these models to viscoelastic hydrodynamics were often limited.\footnote{Indeed, motivated by strongly correlated materials, many such works were primarily interested in more intricate patterns of translational symmetry breaking such that the effective theory is not simply viscoelastic hydrodynamics: see \cite{Baggioli:2014roa,Alberte:2015isw,Alberte:2017cch,Jokela:2017ltu,Andrade:2017ghg,Andrade:2018gqk,Amoretti:2018tzw,Donos:2019tmo,Romero-Bermudez:2019lzz,Donos:2019hpp,Amoretti:2019kuf,Baggioli:2020nay,Andrade:2020hpu,Andrade:2022udb} and \cite{Baggioli:2022pyb} for a review).} Exceptions that studied some basic aspects of the simpler higher-form charged black hole directly are \cite{Grozdanov:2018ewh,Baggioli:2018vfc,Armas:2019sbe,Armas:2020bmo}.

\paragraph{}Our first result is to derive a fluid/gravity duality for small amplitude perturbations of this black hole. That is, we prove explicitly and analytically that the gravitational equations of motion for long wavelength, small amplitude perturbations around this black hole solution are simply the equations of viscoelastic hydrodynamics of \cite{Armas:2019sbe}. At first order in the long wavelength expansion, these hydrodynamic equations ultimately depend on five transport coefficients. For the charged black hole we provide explicit expressions for three of them. Two of these (the conductivity and bulk viscosity) are new, and the other (the thermodynamic pressure) is in agreement with the previous results of \cite{Grozdanov:2018ewh,Armas:2019sbe,Armas:2020bmo}. There are no closed form expressions for the remaining two -- the shear modulus and shear viscosity -- but we show that these are given by solutions to a very simple ordinary differential equation which we evaluate analytically in a variety of limits, as well as numerically. While this result may seem expected, it is worth noting that there have previously been instances where comparisons to holographic models had shown that the then-state-of-the-art theory of viscoelastic hydrodynamics was in fact incomplete.\footnote{To our knowledge, the only time \cite{Armas:2019sbe} was subsequently compared to an explicit model was in the numerical study \cite{Ammon:2020xyv} of the black hole quasinormal modes of a massive gravity-like model at low densities.} Indeed, a partial numerical study of the charged black hole's quasinormal modes \cite{Grozdanov:2018ewh} previously found inconsistencies at high densities in comparison to a hydrodynamic theory predating that of \cite{Armas:2019sbe}.

\paragraph{}Having translated the black hole perturbation theory into the simpler language of viscoelastic hydrodynamics, we then study directly this latter theory, to learn about the former. We show that the equilibrium state has numerous small amplitude, long wavelength instabilities across its parameter space. The state's stability can be enhanced by deforming the theory with a double-trace coupling and we provide an explicit lower bound on this coupling, in terms of the temperature and density, in order that the instabilities are removed from the spectrum. In particular we identify new instabilities than those previously observed in \cite{Grozdanov:2018ewh} from a direct analysis of the Einstein equations. We show that it is one of these new instabilities -- driven by perturbations of an incoherent density -- that is always the first to trigger.

\paragraph{}The black hole with higher-form charge is related by Hodge duality to a black hole spacetime supported by two massless scalar fields with non-trivial spatial profiles. In the scalar formulation, an alternative hydrodynamic description of a subsector of the dynamics has previously been obtained, analytically in the small-charge limit \cite{Bhattacharyya:2008ji,Blake:2015epa} and numerically otherwise \cite{Davison:2014lua}. This alternative description is considerably simpler than viscoelastic hydrodynamics: it is essentially the theory of heat diffusion \cite{Hartnoll:2014lpa,Davison:2014lua} and is insensitive to the value of the double-trace coupling. These two seemingly different descriptions arise due to the non-trivial action of the duality on the hydrodynamic fields (and their sources). Starting from the full theory of viscoelastic hydrodynamics, we show how to derive the heat diffusion theory by implementing this action and subsequently integrating out gapped modes. This yields explicit relations for the transport coefficients in the heat diffusion theory in terms of those of viscoelastic hydrodynamics. Furthermore, we show that the heat diffusion mode that arises has a simple physical interpretation in the original viscoelastic hydrodynamics: it is the collective excitation of the viscoelastic fluid under `no-flux' boundary conditions.

\paragraph{}The structure of the rest of the paper is as follows. In Section \ref{sec:HydroSec} we review the theory of viscoelastic hydrodynamics of \cite{Armas:2019sbe}, exploring in more detail some aspects of linear amplitude perturbations. In Section \ref{sec:holography} we show that this theory, with appropriate values of the transport coefficients, governs the small amplitude, long wavelength fluctuations of an AdS black hole charged under a higher-form gauge potential. We then study in detail this theory of hydrodynamics. In Section \ref{sec:instabilities} we analyse the excitations of the equilibrium state (i.e.~the quasinormal modes of the black hole) and its instabilities, and in Section \ref{sec:hodgeduality} we show how the theory of heat diffusion arises from viscoelastic hydrodynamics. Finally, we conclude in Section \ref{sec:discussion} with a discussion and outlook.

\section{Higher-form hydrodynamics}
\label{sec:HydroSec}

\paragraph{}We start by reviewing aspects of the hydrodynamic theory of relativistic viscoelasticity constructed in \cite{Armas:2019sbe} (see \cite{Delacretaz:2016ivq,Delacretaz:2017zxd,Grozdanov:2018ewh} for earlier work). This has two equivalent formulations: one in terms of higher-form conserved currents, and one in terms of scalar Goldstone fields. We will focus on the homogeneous and isotropic case in (2+1)-dimensions, which is relevant for the holographic theory we still study later.

\paragraph{}Viscoelasticity is the long wavelength, low energy effective theory of crystals. A (2+1)-dimensional crystal lattice can be characterised by the one-form fields $e^I_{\mu}(x^\mu)$ (where $I=1,2$) that are normal to the worldlines of each crystal core. In the continuum limit,  these one-forms are expressed in terms of two scalar fields by
\begin{equation}
    e^I=d\varphi^I.
\end{equation}
This ensures that the one-forms are surface-forming, and the level surfaces of $\varphi^I(x^\mu)$ are the crystal core worldlines. The scalars $\varphi^I$ are called the crystal fields and arise physically as the Goldstone bosons of spontaneously broken translational symmetries. Assuming that the lattice has no topological defects, the crystal fields are single-valued and the one-forms are closed $de^I=0$. From the closed one-forms we can define a pair of two-form currents
\begin{equation}
    J^I=\star e^I,
\end{equation}
that are identically conserved $d\star J^I=0$. These are the currents of the two one-form global symmetries that characterise a viscoelastic crystal.

\subsection{Higher-form viscoelasticity theory}

\paragraph{}Hydrodynamics is a framework for describing the long wavelength, low energy dynamics of thermal systems. The basic assumption is that on these scales the only relevant degrees of freedom are those that are protected from decay by symmetries. In a relativistic fluid with no internal symmetry, these are the energy and momentum densities. The additional degrees of freedom in the hydrodynamic theory of viscoelasticity are the charge densities of the two one-form symmetries. The equations of motion of this theory are then
\begin{equation}
\begin{aligned}
\label{2.3}
	\na_\mu \langle T^\mu_{\;\;\nu}\rangle + \frac{1}{2} \langle J^{I\rho \sig}\rangle (d b_I)_{\nu \rho \sig}  = 0,\quad\quad\quad\quad\quad\quad \na_\mu \langle J^{I\mu \nu}\rangle = 0,
    \end{aligned}
\end{equation}
where $T^{\mu\nu}$ is the energy-momentum tensor and $b_I$ are the external 2-form sources for the 2-form currents $J^I$.\footnote{For consistency with the rest of the paper, in this Section we use the expectation value notation of QFT.}

\paragraph{}To complete the hydrodynamic theory, these equations must be supplemented by constitutive relations between the components of $\langle T^{\mu\nu}\rangle$ and $\langle J^{I\mu\nu}\rangle $. To write these relations it is convenient to express them in terms of auxiliary vector fields $\beta^\mu$ and $\zeta_I^{\mu}$, where $\beta^\mu$ is timelike and $\beta\cdot\zeta_I=0$. Physically, $T=1/\sqrt{-\beta^2}$ is a local temperature, $u^\mu=\beta^\mu/\sqrt{-\beta^2}$ is a local fluid velocity and $\zeta_I^{\mu}$ are local one-form chemical potentials. From these can can construct the spacetime scalar $\gam_{IJ} \equiv \zeta_I \cdot \zeta_J$. The constitutive relations are constructed order-by-order in derivatives of these fields.

\paragraph{}At zeroeth order in derivatives, we obtain the constitutive relations of ideal hydrodynamics. These are 
\begin{equation}
    \begin{aligned}
\label{5.7} 
\langle T_{(0)}^{\mu \nu}\rangle & = \left( s T + q^{IJ} \gam_{IJ} \right) u^\mu u^\nu + p g^{\mu \nu} - q^{IJ} \zeta_I^\mu \zeta_J^\nu,  \quad\quad\quad\quad
\langle J_{(0)}^{I \mu \nu}\rangle & = - 2 q^{IJ} u^{[ \mu} \zeta_J^{\nu ]},
\end{aligned}
\end{equation}
where $p(T,\gamma_{IJ})$ is an arbitrary scalar function and
\be \begin{aligned} \label{6.6}
	s = \frac{\pa p}{\pa T},\quad\quad\quad\quad\quad\quad \quad q^{IJ} = 2 \frac{\pa p}{\pa \gam_{IJ}}.
\end{aligned}  \ee

\paragraph{}At first order in derivatives, the contributions to the constitutive relations are in general very complicated. However, with our assumption of isotropy they simplify somewhat, introducing $\psi^{\mu \nu} = \varep^{\rho \mu \nu} u_{\rho}$ and $\psi^\mu_I=\varepsilon^{\mu\rho\sigma} u_\rho\zeta_{I\sigma}$, to
\be \begin{aligned} \label{EQmatrix}
		\langle T^{\mu \nu}_{(1)}\rangle & = - T\eta^{IJKL}\psi_I^\mu \psi_J^\nu\psi_K^\rho \psi_L^\sig \nabla_{(\rho}\beta_{\sigma)}, \\
		\langle J^{I \mu \nu}_{(1)}\rangle & = T \psi^{\mu \nu}\sigma^{IK}\psi^{\rho \sig} \left(\nabla_{[\rho}\left(\frac{1}{T}\zeta_{K\sigma]}\right)+\frac{1}{2}\beta^\alpha(db_K)_{\alpha\rho\sigma}\right),
\end{aligned} \ee
where $\eta^{IJKL}(T,\gamma_{IJ})$ and $\sigma^{IJ}(T,\gamma_{IJ})$ are the first order transport coefficients and the epsilon tensor is defined such that $\varepsilon_{txy}=\sqrt{-g}$.

\subsection{Small amplitude expansion}
\label{sec:linearisedhydro}

\paragraph{}We are now going to use this hydrodynamic theory to determine how a homogeneous and isotropic equilibrium state responds to a small perturbation. 

\paragraph{}In the equilibrium state, the hydrodynamic fields are independent of the coordinates $x^\mu$. Specifically, it is convenient to choose a frame where $T=\bar{T}$, $u^\mu=\delta ^\mu_0$ and $\zeta^\mu_I=\bar{\gamma}\delta^\mu_I$, where $\bar{T}$ and $\bar{\gamma}$ are constants. The corresponding sources supporting this state are $g_{\mu\nu}=\eta_{\mu\nu}$ and $b_{I\mu\nu}=2\bar{\gamma}\delta^0_{[\mu}\delta_{\nu]I}$. $\bar{\gamma}$ can be thought of as the equilibrium chemical potential of the higher-form charges. We then introduce small perturbations to these sources, parameterised as
\begin{equation}
\label{eq:perturbedsources}
    g_{\mu\nu}=\eta_{\mu\nu}+\delta g_{\mu\nu}(x^\mu),\quad\quad\quad\quad\quad\quad 
    b_{I\mu\nu}=2\bar{\gamma}\delta^0_{[\mu}\delta_{\nu]I}+\delta b_{I\mu\nu}(x^\mu).
\end{equation}
These will induce small perturbations in the hydrodynamic fields, which we parameterise as
\begin{equation}
\begin{aligned}
\label{eq:perturbedhydrofields}
    T(x^\mu) & =\bar{T}+\delta T (x^\mu), \quad\quad
    u^\mu(x^\mu) & =\delta ^\mu_0+\delta u^\mu (x^\mu),\quad\quad
    \zeta^\mu_I(x^\mu)& =\bar{\gamma}\delta^\mu_I+\delta\zeta^\mu_I(x^\mu).
    \end{aligned}
\end{equation}
The profile of the perturbations of the hydrodynamic fields, as a function of the sources, is found by solving the equations of motion \eqref{2.3}. It will be convenient to do this in stages.

\paragraph{}First we use the constitutive relations \eqref{5.7} and \eqref{EQmatrix} to relate perturbations of the stress tensor and charge currents to the perturbations of the hydrodynamic fields. We will work to linear order in perturbations. Assuming that the theory is isotropic, the pressure $p$ cannot depend on $\gamma_{IJ}$ in an arbitrary way but only through its rotational invariants: the determinant and trace. Our equilibrium state is also isotropic: $\bar{\gamma}_{IJ}=\bar{\gamma}^2\delta_{IJ}$ where the bar on the left hand side denotes that the quantity is evaluated in equilibrium. With this information, the general form for the perturbation of the pressure is
\begin{equation}
\begin{aligned}
\label{eq:perturbedp}
    p(T,\gamma_{IJ})&\,=\bar{p}(\bar{T},\bar{\gamma}^2)+\bar{s}\delta T+\frac{1}{2}\bar{Q}\delta^{IJ}\delta\gamma_{IJ}+\frac{1}{2}\frac{\partial\bar{s}}{\partial\bar{T}}\delta T^2+\frac{1}{2}\frac{\partial\bar{Q}}{\partial \bar{T}}\delta^{IJ}\delta T\delta\gamma_{IJ}\\    &\,+\frac{1}{2}\left(\frac{1}{4}\frac{\partial\bar{Q}}{\partial\bar{\gamma}^2}\delta^{IJ}\delta^{KL}+a(\bar{T},\bar{\gamma}^2)\left(2\delta^{I(K}\delta^{L)J}-\delta^{IJ}\delta^{KL}\right)\right)\delta\gamma_{IJ}\delta\gamma_{KL}+\ldots,
    \end{aligned}
\end{equation}
to quadratic order in the perturbations, where $\bar{p}(\bar{T},\bar{\gamma}^2)=p(\bar{T},\bar{\gamma}^2\delta_{IJ})$ is the equilibrium pressure and
\begin{equation}
    \bar{s}(\bar{T},\bar{\gamma}^2)=\frac{\partial\bar{p}}{\partial\bar{T}},\quad\quad\quad\quad\quad\quad \bar{Q}(\bar{T},\bar{\gamma}^2)=\frac{\partial\bar{p}}{\partial\bar{\gamma}^2},
\end{equation}
are the equilibrium entropy and charge densities. The coefficient $a(\bar{T},\bar{\gamma}^2)$ characterises how the pressure responds to variations of the determinant of $\gamma_{IJ}$, keeping its trace (and the temperature) fixed. This information is not contained in derivatives of $\bar{p}$. Using \eqref{6.6}, we obtain the entropy and charge densities to linear order in the perturbations as
\begin{equation}
    \begin{aligned}
    \label{eq:perturbedsq}
    s(T,\gamma_{IJ})=&\,\bar{s}+\frac{\partial\bar{s}}{\partial\bar{T}}\delta T+\frac{1}{2}\frac{\partial\bar{Q}}{\partial\bar{T}}\delta^{IJ}\delta\gamma_{IJ}+\ldots,\\
    q^{IJ}(T,\gamma_{IJ})=&\,\left(\bar{Q}+\frac{\partial\bar{Q}}{\partial\bar{T}}\delta T+\frac{1}{2}\frac{\partial\bar{Q}}{\partial\bar{\gamma}^2}\delta^{KL}\delta\gamma_{KL}\right)\delta^{IJ}\\    &\,+2a(\bar{T},\bar{\gamma}^2)\left(2\delta^{I(K}\delta^{L)J}-\delta^{IJ}\delta^{KL}\right)\delta\gamma_{KL}+\ldots.
    \end{aligned}
\end{equation}
Keeping only terms that will be relevant at leading order in small amplitudes, the first order transport coefficients in \eqref{EQmatrix} can similarly be decomposed as
\begin{equation}
\begin{aligned}
\label{eq:perturbedfirstorder}
    \sigma^{IJ}=&\,\sigma(\bar{T},\bar{\gamma}^2)\delta^{IJ}+\ldots,\\
    \eta^{IJKL}=&\,\frac{\zeta(\bar{T},\bar{\gamma}^2)}{\bar{\gamma}^4}\delta^{IJ}\delta^{KL}+\frac{\eta(\bar{T},\bar{\gamma}^2)}{\bar{\gamma}^4}\left(2\delta^{I(K}\delta^{L)J}-\delta^{IJ}\delta^{KL}\right)+\ldots.
    \end{aligned}
\end{equation}

\paragraph{}Substituting \eqref{eq:perturbedsources}, \eqref{eq:perturbedhydrofields}, \eqref{eq:perturbedp}, \eqref{eq:perturbedsq} and \eqref{eq:perturbedfirstorder} into the constitutive relations \eqref{5.7} and \eqref{EQmatrix} then gives expressions for the stress tensor and charge currents in terms of the hydrodynamic fields and the sources. Substituting these into the equations of motion \eqref{2.3} and solving then tells us the response in terms of the external sources. The hydrodynamic theory completely fixes this response in terms of the five transport coefficients $\bar{p},a,\sigma,\zeta$ and $\eta$. The values of the transport coefficients depend on the microscopic details of the system.

\paragraph{}From now on we will focus on the small amplitude perturbations described above. To reduce notational complexity we will henceforth omit bars on the equilibrium quantities. For example, from now on $s$ denotes the equilibrium entropy density which is a function of the equilibrium temperature $T$ and chemical potential $\gamma$. With this notation, the non-zero components of the energy-momentum tensor and charge currents in equilibrium are
\begin{equation}
\begin{aligned}
\label{eq:eqmexpvalues}
    \langle T^{tt}\rangle =-p+sT+2Q\gamma^2, \quad \langle T^{xx}\rangle=\langle T^{yy}\rangle=p-Q\gamma^2,\quad
    \langle J^{1tx}\rangle = \langle J^{2ty} \rangle=-Q\gamma.
    \end{aligned}
\end{equation}

\paragraph{}To obtain explicit expressions for the small amplitude perturbations, we relabel the coordinates $(x^0,x^1,x^2)$ as $(t,x,y)$ and use rotational symmetry of the state to consider sources that depend only on $(t,x)$. The equations then split into two sets: those involving the transverse perturbations $\{\delta u^y,\delta\zeta_1^y,\delta\zeta_2^x,\delta g_{ty},\delta g_{xy},\delta b_{1ty},\delta b_{1xy},\delta b_{2tx}\}$ and those involving the longitudinal perturbations $\{\delta T,\delta u^x,\delta\zeta_1^x,\delta\zeta_2^y,\delta g_{tt},\delta g_{xx},\delta g_{yy},\delta g_{tx},\delta b_{1tx},\delta b_{2xy},\delta b_{2ty}\}$. The perturbations $\delta u^t$ and $\delta\zeta_I^t$ are fixed in terms of these by the conditions $u\cdot u=-1$ and $\beta\cdot\zeta_I=0$. It will be convenient to work with perturbations of the stress tensor with mixed indices $\langle \delta T^{\mu}_{\;\;\nu}\rangle\equiv \delta\left(\langle T^{\mu\rho}\rangle g_{\rho\nu}\right)$.

\paragraph{}For the transverse perturbations, the relations between the conserved charge densities and the hydrodynamic variables are
\begin{equation}
\begin{aligned}
\label{eq:transversechargeconsrels}
     \langle \delta T^{t}_{\;\; y}\rangle&=(sT+Q\gamma^2)\left(\delta u^y+\delta g_{ty}\right),   \\
     \langle \delta J^{1ty}\rangle&=-Q\delta\zeta_1^y-4a\gamma^2\left(\delta\zeta_1^y+\delta\zeta_2^x+\gamma\delta g_{xy}\right),   \\
    \langle \delta J^{2tx}\rangle&=-Q\delta\zeta_2^x-4a\gamma^2 \left(\delta\zeta_1^y+\delta\zeta_2^x+\gamma\delta g_{xy}\right).   \\
\end{aligned}
\end{equation}
The remaining constitutive relations involving transverse perturbations are
\begin{equation}
\begin{aligned}
\label{eq:transversecurrentconsrels}
     \langle \delta T^{x}_{\;\; y}\rangle&=-\gamma\left(Q+4a\gamma^2 \right)\left(\delta\zeta_1^y+\delta\zeta_2^x+\gamma\delta g_{xy}\right)-\eta\left(\partial_x\delta u^y+\partial_t\delta g_{xy}\right),    \\
     \langle \delta J^{1xy}\rangle&=\gamma Q\delta u^y+\sigma\left(\partial_x\left(\delta \zeta_1^y+\gamma\delta g_{xy}\right)+\delta h_{1txy}\right).
\end{aligned}
\end{equation}

\paragraph{}Similarly, for the longitudinal perturbations the relations between the conserved charge densities and the hydrodynamic variables are 
\begin{equation}
\begin{aligned}
\label{eq:longitudinalchargeconsrels}
     \langle \delta T^{t}_{\;\; t}\rangle=&\,-\left(T\frac{\partial s}{\partial T}+2\gam^2\frac{\partial Q}{\partial T}\right)\delta T-\gamma\left(Q+T\frac{\partial Q}{\partial T}+2\gamma^2\frac{\partial Q}{\partial \gam^2}\right)\left(\delta\zeta_1^x+\delta\zeta_2^y+\gamma\delta g_+\right),   \\
     \langle \delta T^{t}_{\;\; x}\rangle=&\,(sT+Q\gamma^2)\left(\delta u^x+\delta g_{tx}\right),   \\
     \langle \delta J^{1tx}\rangle=&\,-\gamma\frac{\partial Q}{\partial T}\delta T-Q\left(\delta\zeta_1^x+\frac{\gamma}{2}\delta g_{tt}\right)+4a\gamma^2\left(\delta\zeta_2^y-\delta\zeta_1^x+\gamma\delta g_-\right)\\
     &\,-\gamma^2\frac{\partial Q}{\partial\gamma^2}\left(\delta\zeta_1^x+\delta\zeta_2^y+\gamma\delta g_+\right),   \\
    \langle \delta J^{2ty}\rangle=&\,-\gamma\frac{\partial Q}{\partial T}\delta T-Q\left(\delta\zeta_2^y+\frac{\gamma}{2}\delta g_{tt}\right)-4a\gamma^2\left(\delta\zeta_2^y-\delta\zeta_1^x+\gamma\delta g_-\right)\\
    &\,-\gamma^2\frac{\partial Q}{\partial\gamma^2}\left(\delta\zeta_1^x+\delta\zeta_2^y+\gamma\delta g_+\right),   \\
\end{aligned}
\end{equation}
while the remaining constitutive relations are
\begin{equation}
\begin{aligned}
\label{eq:longitudinalcurrentconsrels}
     \langle \delta T_+\rangle&=\left(s-\gamma^2\frac{\partial Q}{\partial T}\right)\delta T-\gamma^3\frac{\partial Q}{\partial\gamma^2}\left(\delta\zeta_1^x+\delta\zeta_2^y+\gamma\delta g_+\right)-\zeta\left(\partial_x\delta u^x+\partial_t\delta g_+\right),   \\
     \langle \delta T_-\rangle&=\gamma\left(Q+4a\gamma^2 \right)\left(\delta\zeta_1^x-\delta\zeta_2^y-\gamma\delta g_-\right)+2\eta\left(\partial_x\delta u^x-\partial_t\delta g_-\right),   \\ 
    \langle \delta J^{2xy}\rangle&=-\gamma Q\delta u^x+\sigma\left(\partial_x\left(\delta\zeta_2^y-\gamma\left(\frac{\delta T}{T}-\delta g_{yy}\right)\right)+\delta h_{2txy}\right),   \\
\end{aligned}
\end{equation}
where we define $\langle \delta T_\pm\rangle=(\langle \delta T^y_{\;\; y}\rangle\pm\langle \delta T^x_{\;\; x}\rangle)/2$ and $\delta g_\pm=(\delta g_{yy}\pm\delta g_{xx})/2$.

\subsection{Scalar formulation of viscoelasticity theory}

\paragraph{}Before explicitly solving the hydrodynamic equations for small amplitude perturbations, we pause to briefly explain the relation between the higher-form formulation and the equivalent Goldstone formulation of viscoelasticity \cite{Armas:2019sbe}. In particular, we explain how our transport coefficients are related to those typically used in the Goldstone formulation. To do this, we temporarily revert to the convention of retaining bars on equilibrium quantities.

\paragraph{}In the Goldstone formulation, the physical distance between crystal cores is characterised by the crystal metric
\begin{equation}
    h^{IJ}=g^{\mu\nu}e^I_{\mu}e^J_{\nu},
\end{equation}
and its inverse $h_{IJ}$ is used to lower $I$ and $J$ indices. In this formulation, the ideal hydrodynamic stress tensor is characterised by a function $P(T,h^{IJ})$ which is related to the pressure $p$ in the higher-form formulation by $P(T,h^{IJ})=p(T,\gamma_{IJ})-q^{IJ}\gamma_{IJ}$ where $h^{IJ}=q^{IK}q^{JL}\gamma_{KL}$. To characterise small amplitude perturbations, we define a reference crystal metric (the crystal metric before external sources are turned on and the system is perturbed) $\mathsf{h}_{IJ}$ and then consider an expansion in small strain $u_{IJ}=\frac{1}{2}(h_{IJ}-\mathsf{h}_{IJ})$. To quadratic order in the strain, this gives
\be \begin{aligned} \label{92}
	P (T,h^{IJ}) = &\, P_f + P_l h^{IJ} u_{IJ} +\frac{1}{2}\left(G - B\right)\left( h^{IJ} u_{IJ} \right)^2+ \left(P_l - G\right) h^{IK} h^{JL} u_{IJ} u_{KL} + \ldots,
\end{aligned} \ee
where $P_f = P (T , \mathsf{h}^{IJ})$. $P_f$ is called the fluid pressure, $P_l$ the lattice pressure, $G$ the shear modulus and $B$ the bulk modulus. They are all functions of the temperature and the reference crystal metric. 

In our case, the system is perturbed away from equilibrium such that 
\begin{equation}
	\mathsf{h}_{IJ} = \bar{h}_{IJ}=\left(\bar{\gamma} \bar{Q}\right)^{-2}\delta_{IJ} \, .
\end{equation}
At this point, we return to the notation of omitting bars on equilibrium quantities. 
Comparing \eqref{92} to \eqref{eq:perturbedp}, we can relate these transport coefficients to those in the higher-form formulation as
\be \label{126} \begin{aligned}
	P_f & = p - 2 Q \gamma^2, \quad P_l = Q \gamma^2, \quad B = Q \gamma^2 \frac{Q + \gamma^2\frac{\partial Q}{\partial\gamma^2}}{Q + 2\gamma^2\frac{\partial Q}{\partial\gamma^2}}, \quad G = Q \gamma^2 \frac{Q + 4 a \gamma^2}{ Q + 8 a \gamma^2}.
\end{aligned} \ee
In other words, our transport coefficient $a$ is morally the shear modulus, while the bulk modulus can manifestly be obtained just from derivatives of the equilibrium pressure $p$.

\paragraph{}At first order in derivatives in the Goldstone formulation, there are three new transport coefficients characterising small amplitude perturbations: the shear viscosity $\eta$, bulk viscosity $\zeta$ and a conductivity. We have defined the two viscosities in the higher-form formulation \eqref{eq:perturbedfirstorder} so that they correspond to those in the Goldstone formulation upon translating. Our conductivity $\sigma$ is related to that in the Goldstone formulation of \cite{Armas:2019sbe} (which we label $\sigma_{G}$) by
\begin{equation}
    \sigma=\left(\frac{Q\gamma sT}{sT+Q\gamma^2}\right)^2\frac{1}{\sigma_G}.
\end{equation}

\subsection{Collective excitations}
\label{sec:hydromodes}

\paragraph{}Finally we will summarise the collective excitations of the viscoelastic fluid. Each mode is characterised by a momentum-space dispersion relation $\omega(k)$, which are the frequencies for which there are non-trivial solutions to the small amplitude hydrodynamic equations in the absence of sources.

\paragraph{}The collective modes of transverse perturbations of the viscoelastic fluid are sound waves. At long wavelengths their dispersion relations are 
\be \begin{aligned}
\label{eq:transversedispers}
	\om = \pm v_\perp k - i \Gam_\perp k^2 + O (k^3), \\
\end{aligned} \ee
where the transverse sound velocity $v_\perp$ and transverse sound attenuation $\Gamma_\perp$ are given by the following combinations of transport coefficients
\be \begin{aligned}
\label{eq:transversedispersion}
	v_\perp^2 & = \frac{G}{sT+Q\gamma^2}, \quad\quad\quad\quad \Gam_\perp & = \frac{1}{2}\left( \frac{\eta}{sT+Q\gamma^2} + \frac{G\sigma}{Q^2\gamma^2} \right).
\end{aligned} \ee

\paragraph{}There are two types of collective modes of the longitudinal perturbations of the viscoelastic fluid: sound waves and a diffusive mode. At long wavelengths the sound waves have dispersion relations
\begin{equation}
\label{eq:sounddispers}
    \omega=\pm v_\parallel k-i\Gamma_\parallel k^2+O(k^3),
\end{equation}
where the sound speed and attenuation are related to the transport coefficients by
\begin{equation}
\begin{aligned}
	v_\parallel^2 &\,= \frac{T\frac{(s -\frac{Q\gam}{2}s_\rho)^2}{c_\rho} +B+G- Q\gamma^2}{ sT  + Q\gamma^2},\\    
	\Gam_\parallel &\,=  \frac{\eta+\zeta}{2(sT + Q\gamma^2)}  + \sigma v_\parallel^2 \frac{(sT + Q\gamma^2)}{2Q^2\gamma^2}  \left( 1 - \frac{s -\frac{Q\gam}{2}s_\rho}{c_\rho v_\parallel^2} \right)^2,
    \end{aligned}
    \end{equation}
and
\begin{equation}
    \begin{aligned}
    \label{eq:wdefns}
	s_\rho = \frac{2\gamma\frac{\pa Q}{\pa T}}{Q+2\gamma^2\frac{\partial Q}{\partial\gamma^2}}, \quad\quad\quad\quad\quad\quad
	c_\rho = T\frac{\pa s}{\pa T} - T\gam s_\rho \frac{\pa Q}{\pa T}.        
    \end{aligned}
\end{equation}
The two quantities in \eqref{eq:wdefns} are expressed in terms of partial derivatives of $s(T,\gamma^2)$ and $Q(T,\gamma^2)$. However, using familiar thermodynamic manipulations they can be expressed more succinctly as
\begin{equation}
\label{crhosrhodefns}
s_\rho=\left(\frac{\partial s}{\partial \left(Q\gam\right)}\right)_T,\quad\quad\quad\quad\quad\quad c_\rho=T\left(\frac{\partial s}{\partial T}\right)_{Q\gam}.
\end{equation}
Recalling from \eqref{eq:eqmexpvalues} that the combination $Q\gam$ is the one-form charge density, $c_\rho$ is simply the heat capacity at constant density while $s_\rho$ is the variation of entropy with charge at fixed temperature.

\paragraph{}At long wavelengths the diffusion mode has the dispersion relation
\begin{equation}
\label{eq:diffusiondispers}
	\omega=-iD_\parallel k^2+O(k^3),
\end{equation}
where the diffusivity is related to the transport coefficients by
\begin{equation}
\label{eq:Dparallelhydro}
    D_\parallel= \frac{1}{c_\rho}\frac{(sT + Q\gamma^2)\sig}{TQ^2\gamma^2} \frac{(B + G-Q\gamma^2)}{v_\parallel^2}.
\end{equation}

\paragraph{}As there are multiple modes in the longitudinal sector, it is instructive to also identify which quantities are transported by each mode. At leading order in the derivative expansion, the longitudinal sound waves \eqref{eq:sounddispers} transport the pressure perturbations $\langle\delta T^{xx}\rangle$ while the diffusive mode \eqref{eq:diffusiondispers} transports the perturbations
\begin{equation}
\label{eq:incoherentdensitydef}
    \delta \rho_{\text{inc}}\equiv\langle \delta J_2^{ty}\rangle+\frac{\gamma Q}{sT+Q\gamma^2}\langle \delta T^{tt}\rangle.
\end{equation}
This is the 1-form analogue of the incoherent density that arises in hydrodynamics with a $0$-form charge \cite{Davison:2015taa}: it diffuses as it carries no momentum (i.e.~no fluid velocity).

\paragraph{}The hydrodynamic modes we have just described are the solutions to the equations of motion under the condition that there are no external sources. An alternative condition is the `no flux' condition $\partial_i\langle\delta J^{2ij}\rangle=0$ for the higher-form currents (and with external sources for the energy-momentum tensor turned off), which due to equation \eqref{eq:longitudinalcurrentconsrels} requires $\delta h_{2txy}\ne0$.\footnote{In hydrodynamics with a $0$-form charge these are sometimes called open circuit boundary conditions (see e.g.~\cite{Davison:2018nxm}): an electric field sets up that prevents current flow.} Physically, the no-flux condition requires the fluid momentum (i.e.~the fluid velocity) to vanish at leading order in derivatives and so the long wavelength transport is diffusive. Specifically, the solution to the hydrodynamic equations in this case has dispersion relation 
\begin{equation}
\label{eq:nofluxdispersion}
    \omega=-iDk^2+O(k^3),
\end{equation}
where
\begin{equation}
\label{eq:nofluxdiffusivity}
    D=\frac{1}{c_\rho}\frac{\left(sT+Q\gamma^2\right)^2\sigma}{T\gamma^2Q^2}.
\end{equation}
Note that $D$ differs from the diffusivity $D_\parallel$ of the source-free hydrodynamic mode \eqref{eq:Dparallelhydro}, as these are two distinct physical processes.

\section{Higher-form fluid/gravity duality}
\label{sec:holography}

\paragraph{}In this Section we will study a planar, asymptotically AdS$_4$ black hole with two one-form symmetries. We will show explicitly that small amplitude perturbations of this spacetime are governed by the theory of hydrodynamics in Section \ref{sec:HydroSec} at sufficiently small frequencies and wavenumbers. We will extract expressions for the thermodynamic and transport coefficients of the fluid governing the dynamics of this black hole. Our results extend the previous studies \cite{Grozdanov:2018ewh,Armas:2019sbe,Armas:2020bmo} of aspects of this black hole.

\paragraph{}The gravitational action is
\begin{equation}
    \label{lag} \begin{aligned} 
 	S&\,=\int d^4 x\sqrt{-G}\left(R+6-\frac{1}{12}\sum_{I=1}^2H^I_{abc}H^{Iabc}\right)+2\int d^3x\sqrt{-\gamma}K+S_{\text{ct}}.
\end{aligned}
\end{equation}
The bulk term in the action is for the metric $G_{ab}$ with Ricci scalar $R$, and two three-form gauge fields $H^I_{abc} = 3 \pa_{[a} B^I_{bc]}=\partial_a B_{bc}^I+\partial_b  B_{ca}^I+\partial_c B_{ab}^I$ dual to two two-form conserved currents $J^{I\mu\nu}$. We will use the usual Poincar\'{e} patch coordinates $(t,x,y,r)$ and cut off the spacetime at a large value $r=\Lambda$ (i.e.~far from the horizon), where the Gibbons-Hawking-York boundary term written above is evaluated with induced metric $\gamma$. We will use indices $a,b,\ldots$ to denote bulk coordinates and indices $\mu,\nu,\ldots$ to denote boundary coordinates.

\paragraph{}The counterterm $S_{\text{ct}}$ is necessary to ensure that the on-shell action remains finite in the limit $\Lambda\rightarrow\infty$ where the cutoff is removed. It has the form \cite{Grozdanov:2018ewh}
\begin{equation}
\label{eq:counterterm}
    S_{\text{ct}}=\int d^3x\sqrt{-\gamma}\left(-4+\sum_{I=1}^2\frac{\Pi^I_{\mu\nu}\Pi^{I\mu\nu}}{4\kappa(\Lambda)}\right),
\end{equation}
where $\Pi^I_{\mu\nu}=n^a H^I_{a\mu\nu}$, with $n^a$ the normal vector to the boundary at $r=\Lambda$. The first term  in \eqref{eq:counterterm} is the usual renormalisation of the stress tensor. In the second term we choose $1/\kappa(\Lambda)=1-\mathcal{M}/\Lambda$. This has two effects. It removes divergences such that the $\Lambda\rightarrow\infty$ limit is well-defined. It also deforms the theory by the relevant double-trace operator $\frac{1}{4}\mathcal{M}\sum_{I=1}^2 J^I_{\mu\nu}J^{I\mu\nu}$ formed from the conserved currents. The double-trace coupling $\mathcal{M}$ has dimensions of energy and will be crucial in what follows. We have chosen this deformation to preserve the isotropy in $IJ$ space of the bulk action, although in principle this can be relaxed. A non-zero coupling $\mathcal{M}$ results in mixed boundary conditions on the bulk fields. We review in more detail the holographic dictionary for the theory \eqref{lag} in Appendix \ref{app:holodictionary}. 

\subsection{Equilibrium thermodynamics}

The action \eqref{lag} has a family of isotropic classical solutions with line element \cite{Bardoux:2012aw}
\begin{equation}
\label{eq:BHmetric}
    ds^2 = \frac{dr^2}{r^2f} + r^2 \left( - f dt^2 + dx^2 + dy^2 \right),\quad\quad\quad\quad f(r)=1 - \frac{\nn^2}{2r^2} - \left( 1 - \frac{\nn^2}{2 r_h^2} \right) \frac{r_h^3}{r^3}, 
\end{equation}
and gauge potential
\begin{equation}
\label{eq:BHBfield}
    B^I_{t a} = m (r - r_h) \delta^I_a,
\end{equation}
with the other components of $B^I_{ab}$ vanishing. $m$ and $r_h>0$ are constants, with the latter being the location of the horizon.

\paragraph{}These are holographically dual to thermal equilibrium states parameterised by the temperature $T$ and chemical potential $\gamma$ for the one-form charges \cite{Grozdanov:2018ewh}. The sources $b_I$ of the one-form charge densities are 
\begin{equation}
b_{I\mu \nu} = \lim_{\Lam \to \infty} \left( B^I_{\mu \nu} - \frac{\Pi^I_{\mu \nu}}{\ka (\Lam)} \right) \Bigg\vert_{r = \Lam}, 
\end{equation}
which correspond to the chemical potential $\gamma=m(\mathcal{M}-r_h)$. Regularity of the Euclidean solution gives the temperature
\begin{equation}
	\label{5.4b} T = \frac{r_h}{4\pi} \left( 3 - \frac{\nn^2}{2 r_h^2} \right)=\frac{r_h}{4\pi} \left( 3 - \frac{\gamma^2}{2 r_h^2(\mathcal{M}-r_h)^2} \right).
\end{equation}
The thermodynamic pressure is given by the on-shell Euclidean action of this solution
\begin{equation}
    \label{7.11} p(T,\gamma^2) = r_h^3 + \frac{\gamma^2\mathcal{M}}{\left(\mathcal{M}-r_h\right)^2} \left(1 - \frac{3r_h}{2\mathcal{M}} \right),
\end{equation}
where $r_h$ should be regarded as a function of $T$, $\gamma^2$ and $\mathcal{M}$ according to equation \eqref{5.4b}. From this we can calculate the thermodynamic derivatives
\begin{equation}
\label{eq:sholodefn}
    s=\frac{\partial p}{\partial T}=4\pi r_h^2,\quad\quad\quad\quad\quad\quad Q=\frac{\partial p}{\partial\gamma^2}=\frac{1}{\mathcal{M}-r_h}.
\end{equation}

\paragraph{}The non-zero expectation values of the stress tensor and charge currents in this equilibrium state are given by \eqref{eq:eqmexpvalues}. In particular, note that the parameter $m$ is simply the density of the one-form charges. Since the equilibrium solution is parameterised more simply in terms of the density $m$ than the chemical potential $\gamma$, we will consider quantities as functions of $T$, $m^2$ and $\mathcal{M}$. In practice we will use the natural bulk variable $r_h$ instead of $T$, where the relation between them is
\begin{equation}
\label{eq:rhdefn}
    r_h(T,m^2)=\frac{2\pi T}{3}+\frac{1}{6}\sqrt{(4\pi T)^2+6m^2}.
\end{equation}

\paragraph{}The other noteworthy feature of the expectation values is that the relevant coupling breaks conformal symmetry and so the stress tensor is not traceless
\begin{equation}
    \langle T^{\mu}_{\;\;\mu}\rangle=-m^2\mathcal{M}(\mathcal{M}-r_h).
\end{equation}
However, we will see below that the fact that this is a double-trace coupling that enters only through the boundary conditions on the fields leads to some simplifications compared to generic conformal symmetry breaking.

\paragraph{}The hydrodynamic constitutive relations in Section \ref{sec:linearisedhydro} depend on second derivatives of the equilibrium pressure. Explicitly, these are
\begin{equation}
\begin{aligned}
   \frac{\partial s}{\partial T}&\,=\frac{(8\pi)^2(\mathcal{M}-r_h)r_h^3}{(\mathcal{M}-r_h)(6r_h^2+m^2)-2m^2r_h} ,\\
    \frac{\partial Q}{\partial T}&\,=\frac{8\pi r_h^2}{(\mathcal{M}-r_h)\left((\mathcal{M}-r_h)(6r_h^2+m^2)-2m^2r_h\right)} ,\\
    \frac{\partial Q}{\partial \gamma^2}&\,=\frac{r_h}{(\mathcal{M}-r_h)^3\left((\mathcal{M}-r_h)(6r_h^2+m^2)-2m^2r_h\right)}.
   \end{aligned}
\end{equation}
From equation \eqref{126}, the bulk modulus of this state is then
\begin{equation}
    B=\nn^2 \left( \mathcal{M} - 2 r_h \frac{\nn^2 + 3 r_h^2}{\nn^2 + 6 r_h^2} \right),
\end{equation}
which agrees with the result in \cite{Armas:2020bmo} after accounting for the factor of two difference in the action. The thermodynamic derivatives defined in \eqref{eq:wdefns} are
\begin{equation}
    s_\rho=\frac{16\pi m r_h^2}{6r_h^2+m^2},\quad\quad\quad\quad\quad\quad c_\rho=8\pi r_h^2\frac{6r_h^2-m^2}{6r_h^2+m^2}.
\end{equation}
Note that for this state
\begin{equation}
\label{eq:shomogeneity}
    c_\rho+ms_\rho=2s.
\end{equation}
This nice relation follows from the fact that it is a double-trace coupling that breaks conformal symmetry. Although in principle physical quantities can depend on three dimensionful parameters $T$, $m$ and $\mathcal{M}$, the entropy density is special. It is given by the Bekenstein-Hawking horizon formula and so can be expressed solely in terms of $T$ and $m$, with no dependence on the coupling $\mathcal{M}$ that enters via asymptotic boundary conditions (this can be seen explicitly by combining equations \eqref{eq:sholodefn} and \eqref{eq:rhdefn}). By dimensional analysis $s$ therefore must have the form $s=T^2S(m/T)$ for some function $S$. Recalling the definitions of $c_\rho$ and $s_\rho$ in terms of partial derivatives of $s$ \eqref{crhosrhodefns}, and that $\rho=-m$, this ensures that \eqref{eq:shomogeneity} is satisfied independent of the form of $S$.

\paragraph{}Finally, for notational convenience it is useful to define the momentum susceptibility as
\begin{equation}
    \chi\equiv sT+Q\gamma^2,
\end{equation}
in what follows. 

\subsection{Black hole perturbations}

\paragraph{}We are now going to study low frequency, long wavelength perturbations of the equilibrium black brane. We will work to linear order in the amplitudes of these perturbations, which we denote as $\delta G_{ab}$ and $\delta H^I_{abc}=3\partial_{[a}\delta B^I_{bc]}$. 

\paragraph{}The equations of motion for these fields are invariant under local diffeomorphisms $\xi^a$ and 1-form gauge transformations $\lambda^I_a$. It will be helpful to incorporate this directly from the outset. The metric and gauge potential perturbations transform as
\begin{equation}
\begin{aligned}
\del G_{ab} \rightarrow \del G_{ab}+ 2 \nabla_{( a} \xi_{b )},\quad\quad\quad
\del B^I_{ab} \rightarrow \del B^I_{ab} + \xi^d \nabla_d B^I_{ab} + 2 B^I_{c [b} \nabla_{a ]} \xi^c + 2 \nabla_{[a} \lam^I_{b]},
\end{aligned}
\end{equation}
where $\nabla$ is the covariant derivative of the equilibrium metric \eqref{eq:BHmetric} and the potential on the right hand side is that of the equilibrium solution \eqref{eq:BHBfield}. The corresponding transformation of the field strength perturbation is
\begin{equation}
\delta H^I_{abc} \rightarrow \delta H^I_{abc} + \xi^d \nabla_d H^I_{abc} + 3 H^I_{d [bc} \nabla_{a ]} \xi^d,
\end{equation}
where the field strength on the right hand side is that of the equilibrium solution.

\paragraph{}With this in mind, we will parameterise our perturbations by
\begin{equation}
\label{2.8} \begin{aligned}
	\del G_{\mu \nu}  &\,= \hat{g}_{\mu \nu} - 2 \pa_{(\mu} \xi_{\nu)} - \xi^r \pa_r G_{\mu \nu}, \\
 	\del G_{ar}  &\,= - 2 \pa_{(a} \xi_{r)} - \xi^r \pa_r G_{ar}, \\
	\del H^I_{abc} &\,= \hat{h}^I_{abc} - \nn \ep^{IJ} \hat{\ep}_{abcd} \left( \del_J^d \pa_e \xi^e - \pa_J \xi^d \right),\\
\end{aligned}
\end{equation}
where the metric $G$ on the right hand side is the background metric and the epsilon symbol defined such that $\hat{\ep}_{rtxy} = 1$. The fields $\hat{g}_{\mu\nu}$ are simply the metric perturbations in radial gauge and the equation \eqref{2.8} tells us that the general solution is found by solving in this gauge and then performing an arbitrary diffeomorphism. This parameterisation is useful since only the fields $\hat{g}_{\mu\nu}$ and $\hat{h}^I_{abc}$ appears in the equations of motion.  We define $\hat{h}^I_{abc}=3\partial_{[a}\hat{b}^I_{bc]}$ with $\hat{b}^I_{ra}=0$ i.e.~we work in radial gauge for the gauge potentials.
Finally, we Fourier transform these fields as follows
\begin{equation}
\begin{aligned}
\label{eq:fouriertransconvens}
    \hat{g}_{\mu\nu}=\int\frac{d\omega dk}{(2\pi)^2}\hat{g}_{\mu\nu}(r,\omega,k)e^{-i\omega t+ikx},
    \end{aligned}
\end{equation}
with an analogous expression for perturbations of the gauge fields. We have used rotational symmetry of the solution in the $(x,y)$ plane to align the wavevector $\vec{k}$ with the $x$ direction without loss of generality. From now on we raise and lower internal $I,J$ indices with the Kronecker delta, and raise and lower spacetime indices on $\hat{g}$, $\hat{h}$ and $\hat{b}$ with the equilibrium metric \eqref{eq:BHmetric}.

\paragraph{}In total there are ten components of the Einstein equations
\begin{equation}
\label{eq:EinsteinEqsGeneral}
    R^{ab} - \left( \frac{R}{2} + 3 \right) G^{ab} + \frac{1}{4} \left( \frac{G^{ab} G^{cd}}{6} - G^{ac} G^{bd} \right) G^{em} G^{fn} \sum_{I=1}^2 H^I_{cef} H^I_{dmn}=0,
\end{equation}
and twelve components of the Maxwell$_I$ equations
\begin{equation}
    \frac{1}{2} \pa_c \left( \sqrt{-G} H^I_{def} G^{dc} G^{ea} G^{fb} \right)=0,
\end{equation}
that must be satisfied. Ultimately we will impose all of these, but it will be convenient to do this in stages. We will first find the general solutions of all of the equations of motion except the $r\mu$ components of the Einstein equations and the $ry$ components of the Maxwell$_I$ equations. We will refer to these as the `bulk equations'. The remaining equations are all radial constraints and will be imposed later.

\paragraph{}The bulk equations can be further separated into two sets. The first set are those involving the fields $\hat{g}_{xy}$, $\hat{g}_{ty}$, $\hat{b}^1_{ty}$, $\hat{b}^1_{xy}$ and $\hat{b}^2_{tx}$. We call these the transverse bulk equations. The second set are those involving the remaining fields $\hat{g}_{xx}$, $\hat{g}_{yy}$ $\hat{g}_{tt}$, $\hat{g}_{xt}$, $\hat{b}^1_{tx}$, $\hat{b}^2_{ty}$ and $\hat{b}^2_{xy}$ and we call these the longitudinal bulk equations. This mirrors the decomposition we saw in the theory of hydrodynamics in Section \ref{sec:HydroSec}. These two sets of equations decouple from one another and so we can solve them separately.

\subsubsection{Transverse bulk equations}

\paragraph{}We start with the transverse bulk equations for $\hat{g}_{xy}$, $\hat{g}_{ty}$, $\hat{b}^1_{ty}$, $\hat{b}^1_{xy}$ and $\hat{b}^2_{tx}$. We want to solve these in a low frequency, long wavelength expansion. More precisely, we take $\omega\sim k\sim\epsilon\ll 1$, keeping $\hat{g}_{\mu\nu}$ and $\hat{b}_{I\mu\nu}$ fixed and will neglect terms of order $\epsilon^2$ and higher in the equations of motion. The general solutions to these equations depend on numerous integration constants (functions of only $t$ and $x$). When writing the solutions below, we always use the holographic dictionary in Appendix \ref{app:holodictionary} to label the integration constants in terms of dual field theory operator expectation values $\langle\delta T^{\mu}_{\;\;\nu}\rangle\equiv \delta\left(\langle T^{\mu\rho}\rangle g_{\rho\nu}\right)$, $\langle\delta J_I^{\mu\nu}\rangle$ and sources $\delta g_{\mu\nu}$, $\delta b_{I\mu\nu}$.

\paragraph{}The first transverse bulk equations are the $tx$, $rx$ and $rt$ components of the Maxwell$_2$ equation
\begin{equation}
\label{eq:transeom1}
    \partial_a\left(r^2\hat{h}_2^{rtx}+m\hat{g}^x_y\right)=0,
\end{equation}
where $a=\{r,t,x\}$. The $r$-component of \eqref{eq:transeom1} is trivial to integrate, yielding
\begin{equation}
\label{eq:transeom2}
    r^2\hat{h}_2^{rtx}+m\hat{g}^x_y=\braket{\delta J_2^{tx}}.
\end{equation}
The quantity on the right hand side is the $r$-independent integration constant.

\paragraph{}The $t$ and $x$ components of \eqref{eq:transeom1} are constraints and imposing them requires that
\begin{equation}
\label{eq:Jconstraint1}
\partial_t\braket{\delta J_2^{tx}}=\partial_x\braket{\delta J_2^{tx}}=0.    
\end{equation}
We will impose these conditions from now on.

\paragraph{}Solving the rest of the transverse bulk equations is more challenging. The $ty$ and $xy$ components of the Einstein equations can be written
\begin{equation}
    \begin{aligned}
 \label {eq:transversefluceqA} \pa_r \left( r^4 \pa_r\hat{g}^y_t \right) - \nn \hat{h}^1_{rxy}=0, \quad\quad\quad
 \pa_r \left( r^4f \pa_r\hat{g}^x_y \right) + \nn \hat{h}^1_{try} =-\nn \braket{\delta J_2^{tx}},        
    \end{aligned}
\end{equation}
while the $ty$ and $xy$ components of the Maxwell$_{1}$ equations are
\begin{equation}
    \begin{aligned}
     \label {eq:transversefluceqB} \pa_r \left( r^2 \hat{h}_1^{try} - \nn \hat{g}^x_y\right)=0, \quad\quad\quad\quad\quad\quad
\pa_r \left( r^2 \hat{h}_1^{rxy} - \nn \hat{g}^y_t\right)=0.
\end{aligned}
\end{equation} 

\paragraph{}To find the solutions, we first algebraically solve the first equations in \eqref{eq:transversefluceqA} and \eqref{eq:transversefluceqB} for $\hat{h}^1_{rxy}$ and $\pa_r\hat{g}^x_y$ respectively. Substituting these solutions into the second equations in \eqref{eq:transversefluceqA} and \eqref{eq:transversefluceqB}, and then simplifying, gives the decoupled second order equations
\begin{equation}
\begin{aligned}
   	\label{eq:transverseeom}
    &\,\pa_r \left( r^8f{f'}^2  \pa_r \left( \frac{\pa_r\hat{g}^y_t}{f'} \right) \right)  = 0, \\
	&\,\pa_r \left( r^4f \pa_r \left( r^2 \hat{h}_1^{try} \right) \right) - \nn^2 r^2 \hat{h}_1^{try} = - \nn^2 \braket{\delta J_2^{tx}}, 
    \end{aligned}
\end{equation}
where prime denotes a derivative with respect to $r$.

\paragraph{}The first of these can be formally integrated for $\pa_r\hat{g}^y_t$. It is useful to define
\begin{equation}
\label{eq:Phiindefdefn}
    \Phi(r)=r^4f'\int\frac{dr}{r^8f{f'}^2}.
\end{equation}
This indefinite integral is only defined up to a constant. We choose this constant such that the integral has no $r$-independent term when expanded as $r\rightarrow\infty$. With this choice, the general solution is (again, after appropriately labelling the integration constants)
\be \begin{aligned} \label{27a}
	\pa_r\hat{g}^y_t & = \left(\delta g_{ty} + \frac{\braket{\del J_1^{xy}}}{m} \right) \frac{f'}{\nn} - \left( \braket{\delta T^t_{\;\; y}} - \chi \left(\delta g_{ty} + \frac{\braket{\del J_1^{xy}}}{m}\right) \right) \frac{m^2}{r^4}\Phi(r).
\end{aligned} \ee
We will not need to evaluate the integral $\Phi(r)$ explicitly.

\paragraph{}The second equation in \eqref{eq:transverseeom} is even less tractable. We denote the two linearly independent solutions to the homogeneous equation for $r^2\hat{h}_1^{try}$ as $\Theta_1(r)$ and $\Theta_2(r)$. Using the Wronskian method, we can express the second solution in terms of the first by
\begin{equation}
 	\Theta_2 (r) = \Theta_1 (r) \int \frac{dr}{r^4f \Theta_1 (r)^2},  
\end{equation}
where the constant in the indefinite integral is chosen as described below equation \eqref{eq:Phiindefdefn}. A local analysis indicates that one of the solutions is non-zero (and finite) on the horizon. Our convention is to call this solution $\Theta_1(r)$ (i.e.~$\Theta_1(r_h)\ne0$). A local analysis near the boundary reveals that
\begin{align} \label{3.25}
	\Theta_1 (r) = \Theta_1(\infty)\left(1 - \frac{\nn^2}{2 r^2} - \frac{\nn^2 \theta(r_h,m^2)}{3 r^3} + \ldots\right),
\end{align}
where $\Theta_1(\infty)$ is an overall normalisation and the specific form of $\theta(r_h,m^2)$ can only be determined by explicitly solving the equation. We will be able to show that low frequency, long wavelength perturbations of the black hole obey the hydrodynamic equations of Section \ref{sec:HydroSec} without doing this. However, an explicit solution is required to obtain an explicit expression for the shear modulus and shear viscosity of the fluid and we will return to this in Section \ref{sec:Thetasols} below. For now, we use linearity of the equation of motion to set $\Theta_1(\infty)=1$ so that the general solution to the second equation in \eqref{eq:transverseeom} is then
\begin{equation}
\begin{aligned}
\label{eq:h1trysol}
    	r^2 \hat{h}_1^{try} = \braket{\delta J_2^{tx}} & + \left( \nn \delta g_{xy} - \braket{\del J_2^{tx}} - \braket{\del J_1^{ty}} \right) \Theta_1 (r) \\
        & - m\left( \braket{\delta T^x_{\;\; y}} +m\left(\mathcal{M}+\theta\right) ( m\delta g_{xy} - \braket{\del J_2^{tx}} - \braket{\del J_1^{ty}} ) \right) \Theta_2 (r).
        \end{aligned}
\end{equation}

\paragraph{}We are now in a position to obtain expressions for the metric and gauge potentials. Integrating the first equations in \eqref{eq:transversefluceqA} and \eqref{eq:transversefluceqB} respectively give
\begin{equation}
\begin{aligned}
\label{eq:transversefinalsols1}
\hat{b}^1_{xy} = \frac{r^4 \pa_r\hat{g}^y_t - \braket{\delta T^t_{\;\; y}}}{\nn} + \delta b_{xy}^1, \quad\quad\quad\quad\quad\quad \hat{g}^x_y = \frac{r^2 \hat{h}_1^{try} + \braket{\delta J_1^{ty}}}{\nn},
\end{aligned}
\end{equation}
where the fields on the right hand side are given in \eqref{27a} and \eqref{eq:h1trysol} above. Integrating the second equations in \eqref{eq:transversefluceqA} and \eqref{eq:transversefluceqB} respectively then give 
\begin{equation}
    \begin{aligned}
    \label{eq:transversefinalsols2}
     \hat{b}^1_{ty} &= \frac{r^4f \pa_r\hat{g}^x_y + \braket{\delta T^x_{\;\; y}}}{\nn} + \delta b_{ty}^1 + (r - \mathcal{M}) \braket{\delta J_2^{tx}}, \\     
    \hat{g}^y_t &= \frac{r^2 \hat{h}_1^{rxy} - \braket{\delta J_1^{xy}}}{\nn},
    \end{aligned}
\end{equation}
where the fields on the right hand side are calculated from \eqref{eq:transversefinalsols1}. Finally, integrating 
\eqref{eq:transeom2} gives $\hat{b}^2_{tx}$, but we will not need its explicit form.

\subsubsection{Longitudinal bulk equations}

\paragraph{}We now move to the longitudinal bulk equations for $\hat{g}_{xx}$, $\hat{g}_{yy}$ $\hat{g}_{tt}$, $\hat{g}_{xt}$, $\hat{b}^1_{tx}$, $\hat{b}^2_{ty}$ and $\hat{b}^2_{xy}$, which we solve in the same low frequency and long wavelength expansion described before. It will be convenient to define $\hat{g}_{\pm}=(\hat{g}^y_y \pm \hat{g}^x_x)/2$. As above, when solving we always relabel integration constants (functions of only $t$ and $x$) in terms of dual field theory operator expectation values $\langle\delta T^{\mu}_{\;\;\nu}\rangle$, $\langle\delta J_I^{\mu\nu}\rangle$ and sources $\delta g_{\mu\nu}$, $\delta b_{I\mu\nu}$ using the holographic dictionary in Appendix \ref{app:holodictionary}. We also use the following trace Ward identity for small amplitude perturbations to simplify expressions where appropriate
\begin{equation}
\label{eq:traceWardidentityperturbed}
    \langle\delta T^t_{\;\; t}\rangle+2\langle \delta T_+\rangle=m^2\mathcal{M}\left(-\delta g_+ +\delta g_{tt}+\frac{\langle\delta J_1^{tx}\rangle+\langle\delta J_2^{ty}\rangle}{m}\right).
\end{equation}
This identity follows from the equations of motion and the holographic dictionary and the non-zero right hand side reflects the breaking of conformal symmetry by the double-trace coupling $\mathcal{M}$.
 
\paragraph{}The simplest of these equations are the $tx$, $rx$ and $rt$ components of the Maxwell$_1$ equation
\begin{equation}
	\label{3.6} \pa_{a}\left(r^2\hat{h}_1^{rtx} + \frac{\nn}{2} \left( \hat{g}_t^t-2\hat{g}_- \right)\right)=0,
\end{equation}
where $a=\{r,t,x\}$. The $r$-component of this can be straightforwardly integrated yielding
\begin{equation}
\label{eq:h1rtxsol}
   r^2\hat{h}_1^{rtx} + \frac{\nn}{2} \left( \hat{g}_t^t -2\hat{g}_- \right)=\braket{\delta J_1^{tx}} - \nn \frac{\delta g_\sig^\sig}{2}.
\end{equation}
The remaining components of this equation are constraints and imposing them, as we do from now on, requires that 
\begin{equation}
\label{eq:Jconstraint2}
\partial_t\left(\braket{\delta J_1^{tx}} - \frac{m}{2}\delta g^\sigma_\sigma \right)=\partial_x\left(\braket{\delta J_1^{tx}} - \frac{m}{2}\delta g^\sigma_\sigma \right)=0.   
\end{equation}

\paragraph{}The next simplest are the $ty$ and $xy$ components of the Maxwell$_2$ equation
\begin{equation}
\begin{aligned}
\label{3.1b} 
\pa_r \left( r^2 \hat{h}_2^{rty} + \frac{\nn}{2} \left( \hat{g}_t^t +2\hat{g}_- \right) \right)=0,  \quad\quad\quad\quad\quad \pa_r \left( r^2 \hat{h}_2^{rxy} + \nn \hat{g}_t^x \right)=0,
\end{aligned} 
\end{equation}
which upon integration give
\begin{equation}
    \begin{aligned}
\label{2final}
r^2 \hat{h}_2^{rty} + \frac{\nn}{2} \left( \hat{g}_t^t + 2\hat{g}_- \right)&= \braket{\delta J_2^{ty}} - \nn \frac{\delta g_\sig^\sig}{2}, \\
r^2 \hat{h}_2^{rxy} + \nn \hat{g}_t^x &= \braket{\delta J_2^{xy}}.        
    \end{aligned}
\end{equation}

\paragraph{}Now we turn to the Einstein equations. We start with the $tx$ component
\begin{equation}
    			\label{9anew} f \pa_r \left( r^4 \pa_r \hat{g}^x_t \right) + \nn r^2 \hat{h}_2^{rxy} =0.
\end{equation}
Solving this for $\hat{h}_2^{rxy}$, substituting into the second equation in \eqref{3.1b} and simplifying gives
\begin{equation}
	\pa_r \left( r^8f{f'}^2 \pa_r \left( \frac{\partial_r\hat{g}^x_t}{f'} \right) \right) = 0.
\end{equation}
This has the same functional form as the first equation in \eqref{eq:transverseeom} and so integrating it gives
\be \begin{aligned}
	\partial_r\hat{g}^x_t & = \left( \delta g_{xt} - \frac{\braket{\del J_2^{xy}}}{\nn} \right) f'  -  \left( \braket{\delta T^t_{\;\; x}} - \chi\left( \delta g_{tx} - \frac{\braket{\del J_2^{xy}}}{m}\right) \right) \frac{m^2\Phi(r)}{r^4}.
\end{aligned} \ee
Substituting this into equation \eqref{9anew} gives
\be
	\hat{h}_2^{rxy}  = - \nn \frac{f}{r^2} \left( \left( \delta g_{tx} - \frac{\braket{\del J_2^{xy}}}{\nn} \right) + \left( \braket{\delta T^t_{\;\; x}} - \chi\left(\delta g_{tx} - \frac{\braket{\del J_2^{xy}}}{m}\right) \right) \int \frac{dr}{r^4f^2} \right),
\ee
after an integration by parts using $\partial_r(r^4f')=m^2$. The constant is chosen such that the indefinite integral has no $r$-independent term when expanded at large $r$. Using the second equation in \eqref{2final}, we then obtain
\begin{equation}
    \hat{g}_t^x = \frac{\braket{\delta J_2^{xy}}}{\nn} + f \left( \left( \delta g_{tx} - \frac{\braket{\del J_2^{xy}}}{\nn} \right) + \left( \braket{\delta T^t_{\;\; x}} - \chi\left( \delta g_{tx} - \frac{\braket{\del J_2^{xy}}}{m}\right) \right) \int \frac{dr}{r^4f^2} \right).
\end{equation}

\paragraph{}The next simplest Einstein equation is the difference between the $xx$ and $yy$ components. After substituting in the solutions \eqref{eq:h1rtxsol} and the first equation in \eqref{2final} for $\hat{h}_1^{rtx}$ and $\hat{h}_2^{rty}$ it is 
\be \begin{aligned} \label{3.20}
	\pa_r \left( r^4f \pa_r \hat{g}_- \right) - \nn^2  \hat{g}_- = \nn \left( \braket{\delta J_1^{tx}} - \braket{\delta J_2^{ty}} \right).
\end{aligned} \ee
This has the same functional form as the second equation in \eqref{eq:transverseeom} and the general solution is
\be \begin{aligned}
	\hat{g}_-  = \frac{\braket{\delta J_2^{ty}} - \braket{\delta J_1^{tx}}}{\nn} & + \left( \delta g_- + \frac{\braket{\del J_1^{tx}} - \braket{\del J_2^{ty}}}{\nn} \right) \Theta_1 (r) \\
    &- \left( \braket{\delta T_-} +m \left(\mathcal{M}+\theta\right) (m\delta g_- + \braket{\del J_1^{tx}} -\braket{\del J_2^{ty}} ) \right) \Theta_2 (r).
\end{aligned} \ee

\paragraph{}Finally, we have the $tt$, $rr$ and the sum of the $xx$ and $yy$ components of the Einstein equation. These are equations for the fields $\hat{g}^t_t$ and $\hat{g}_+$. The $tt$ component can be written
\be \begin{aligned} \label{125new}
	\pa_r \left( f^{3/2} r^4 \pa_r \left( \frac{\hat{g}_+ + \frac{\braket{\delta J_1^{tx}} + \braket{\delta J_2^{ty}}}{\nn} -\delta g^\sigma_\sigma}{\sqrt{f}} \right) \right) = 0.
\end{aligned} \ee
Integrating this gives
\be \begin{aligned} \label{13}
	\hat{g}_+ =\delta g^\sigma_\sigma- \frac{\braket{\delta J_1^{tx}} + \braket{\delta J_2^{ty}}}{\nn} & +  \sqrt{f} \bigg( - \delta g_+ + \delta g_{tt} + \frac{\braket{\del J_1^{tx}} + \braket{\del J_2^{ty}}}{\nn} \\
    & + \frac{\braket{\delta T^t_{\;\; t}} - \chi\left( \frac{\braket{\del J_1^{tx}} + \braket{\del J_2^{ty}}}{m} +\delta g_{tt} - \delta g_+\right)}{2} \int \frac{dr}{ r^4f^{3/2}} \bigg).
\end{aligned} \ee
To obtain $\hat{g}^t_t$, we use the $rr$ component of the Einstein equations, which is first order in radial derivatives. Combining this with \eqref{125new} yields
\be \begin{aligned} \label{127}
	\pa_r \hat{g}^t_t = \frac{1}{r} \pa_r \left( r^2 \pa_r \left(\hat{g}_+ + \frac{\braket{\delta J_1^{tx}} + \braket{\delta J_2^{ty}}}{\nn}-\delta g^\sigma_\sigma\right) \right),
\end{aligned} \ee
and integrating by parts then gives
\be \begin{aligned} \label{3.32}
	\hat{g}^t_t & = - \delta g_{tt} - \delta g_+  + \pa_r \left( r \hat{g}_+ \right).
\end{aligned} \ee
The sum of the $xx$ and $yy$ components of the Einstein equation is implied by the equations we have already solved above and so does not need to be considered separately.

\paragraph{}The final step is to obtain the gauge potentials from the results above. This is done by integrating \eqref{9anew}, the first equation in \eqref{2final} and \eqref{eq:h1rtxsol} respectively for $\hat{b}^2_{xy}$, $\hat{b}^2_{ty}$ and $\hat{b}^1_{tx}$. This gives
\begin{equation}
    \begin{aligned}
    \hat{b}^2_{xy} = &\,\frac{\braket{\delta T^t_{\;\; x}} - r^4 \pa_r \hat{g}^x_t}{\nn} + \delta b_{xy}^2,\\
    \hat{b}^2_{ty} =&\,\delta b^2_{ty} - \mathcal{M} \left( \nn \left( \delta g_{yy}-\delta g_{tt} \right) - \braket{\delta J_2^{ty}} \right) +mr\left(\frac{\hat{g}_+}{2}-\delta g_{tt}+\frac{\delta g_+}{2}-\frac{\langle\delta J_1^{tx}\rangle}{m}\right)\\
    &\,+\left(m\delta g_-+\langle\delta J_1^{tx}\rangle-\langle J_2^{ty}\rangle\right)\left((\mathcal{M}+\theta)\left(1-r^4f\partial_r\Theta_2\right)+\left(\frac{r^4f}{m^2}\partial_r\Theta_1-\theta\right)\right)\\
    &\,+\frac{\langle\delta T_-\rangle}{m}\left(1-r^4f\partial_r\Theta_2\right),
    \end{aligned}
\end{equation}
where the constant in the indefinite integral is chosen as described below equation \eqref{eq:Phiindefdefn}. We will not need the explicit form of $\hat{b}^1_{tx}$.

\subsection{Conservation equations}

\paragraph{}We have just solved 17 of the 22 equations of motion. The solutions are parameterised by integration constants that we have labelled in terms of thermal expectation values and sources of field theory operators. We are now going to impose the remaining 5 equations. These are all radial constraints. They require the field theory expectation values and sources to obey local conservation equations for energy, momentum and one-form charges.

\paragraph{}The remaining equations involving the transverse fields are the $ry$ component of the Maxwell$_1$ equation and the $ry$ component of the Einstein equation
\begin{equation}
\begin{aligned}
\label{1.21Trans}
&  r^2 \pa_t \hat{h}_1^{try} - r^2 \pa_x \hat{h}_1^{rxy} - \nn\left(\pa_t\hat{g}^x_y-\pa_x\hat{g}^y_t\right)=0, \\
& \pa_r\pa_t\hat{g}^y_t -f \pa_r\pa_x\hat{g}^x_y - \frac{\nn}{r^4} \hat{h}^1_{txy}=0.
\end{aligned}
\end{equation}
Inserting the solutions for the metric and gauge field perturbations from the last Section, these reduce to
\begin{equation}
    \begin{aligned}
    \label{eq:3.52}
			&\pa_t \braket{\del J_1^{ty}} + \pa_x \braket{\del J_1^{xy}}=0 , \\
            & \pa_t \braket{\del T^t_{\;\; y}} + \pa_x \braket{\del  T^x_{\;\; y}} - \nn \delta h_{txy}^1=0 
    \end{aligned}
\end{equation}

\paragraph{}The remaining equations that involve the longitudinal fields are the $ry$ component of the Maxwell$_2$ equation, and the $rx$ and $rt$ components of the Einstein equations 
\begin{equation}
\begin{aligned}
\label{1.21Long} 
 & \pa_t \left[  2 r^2 \hat{h}_2^{rty} + \nn \left( \hat{g}_t^t + \hat{g}_y^y - \hat{g}_x^x \right) \right] + 2 \pa_x \left( r^2 \hat{h}_2^{rxy} + \nn \hat{g}_t^x \right)=0,\\
&  \nn r^2 \hat{h}_2^{txy} -\frac{\partial_t\partial_r\hat{g}^x_t}{f}- \pa_x \partial_r(\hat{g}^t_t+\hat{g}^y_y) - \frac{f'}{2f} \pa_x \hat{g}^t_t=0,\\
&-f\pa_t\pa_r\left(\hat{g}^x_x+\hat{g}^y_y\right) +f \pa_x\pa_r\hat{g}^x_t + \frac{f'}{2} \pa_t \left( \hat{g}^x_x + \hat{g}^y_y \right)- f' \pa_x \hat{g}^x_t=0.   
\end{aligned}
\end{equation}
Inserting the solutions for the metric and gauge field perturbations from the last Section, these further require that
\begin{equation}
    \begin{aligned}
\label{eq:3.54} 
			& \pa_t \left(\braket{\del J_2^{ty}}-\frac{m}{2}\delta g^\sigma_\sigma\right) + \pa_x \braket{\del J_2^{xy}}=0 , \\
            & \pa_t \braket{\del T^t_{\;\; x}} + \pa_x \braket{\del T^x_{\;\; x}} + \nn \delta h_{txy}^2 - \frac{\chi}{2} \pa_x \delta g_{tt}=0 , \\
            & \pa_t \braket{\del T^t_{\;\; t}} + \pa_x \braket{\del T^x_{\;\; t}} - \chi\left(\pa_t\delta g_+-\partial_x\delta g_{tx}\right)=0.        
    \end{aligned}
\end{equation}

\paragraph{}The five conditions in \eqref{eq:3.52} and \eqref{eq:3.54}, and the four conditions in \eqref{eq:Jconstraint1} and \eqref{eq:Jconstraint2} are simply the local conservation laws $\na_\mu \braket{T^\mu_{\;\;\nu}} + \frac{1}{2} (d b_I)_{\nu \rho \sig} \braket{J^{I\rho \sig}} = 0$ and $\nabla_\mu \braket { J^{I\mu \nu}}=0$ evaluated for linearised perturbations around the equilibrium state. These are the hydrodynamic equations of motion.

\subsection{Constitutive relations}
\label{sec:consrelationshigherform}

\paragraph{}Having solved all of the gravitational equations of motion, the last task is to impose boundary conditions on these solutions. Specifically, we demand that the solutions are regular in ingoing Eddington-Finkelstein-like coordinates. This will give us four new conditions on the integration constants of the solution above, corresponding to the hydrodynamic constitutive relations. Combined with the conservation equations we derived before, we have therefore derived the theory of hydrodynamics governing the long wavelength perturbations of the black hole.

\subsubsection{Regularity conditions}

\paragraph{}To obtain the causal response of the field theory, we first change from the $t$ coordinate to the ingoing null coordinate $v$ defined as
\begin{equation}
\label{eq:nullcoorddefn}
    dv=dt+\frac{dr}{r^2f}.
\end{equation}
After Fourier transforming, the relation between the field perturbations $\{\delta G_{ab},\delta H_{abc}^I\}$ above and those in the ingoing coordinate system $\{\delta \mathcal{G}_{AB},\delta\mathcal{H}_{ABC}^I\}$ are
\begin{equation}
    \begin{aligned}
    \label{eq:NHexpmunu}
        \delta G_{\{tt,ti,ij\}}&\,=e^{i\omega (t-v)}\delta\mathcal{G}_{\{vv,vi,ij\}},\quad\quad\quad\quad\quad\quad
        \delta H_{txy}^I=e^{i\omega(t-v)}\delta\mathcal{H}_{vxy}^I,\\
                \delta H_{rxy}^I-\frac{\delta H_{txy}^I}{r^2f}&\,=e^{i\omega(t-v)}\delta \mathcal{H}_{rxy}^I,\quad\quad\quad\quad\quad\quad\quad\quad \delta H_{rti}^I=e^{i\omega(t-v)}\delta\mathcal{H}_{rvi}^I,
    \end{aligned}
\end{equation}
where the indices $i$ and $j$ are for $\left\{ x,y\right\}$. 

\paragraph{}Near-horizon regularity of the terms on the right hand side of equation \eqref{eq:NHexpmunu} imposes relations on the near-horizon expansions of the terms on the left hand side. Denoting the near-horizon expansions of the combinations on the left hand side of \eqref{eq:NHexpmunu} as
\begin{equation}
\label{eq:genericNHexpn}
    Z(r\rightarrow r_h)=\Gamma[Z]\log f+\alpha[Z]+\ldots,
\end{equation}
where $\alpha[Z]$ and $\Gamma[Z]$ are $r$-independent quantities that depend on the particular field $Z$, the regularity conditions require that
\begin{equation}
\label{eq:genericNHBC}
    \Gamma[Z]=-\frac{i\omega}{4\pi T}\alpha[Z],
\end{equation}
to the order in the derivative expansion that we are working.

\paragraph{}In addition, the radial components of the metric perturbations in the two coordinate systems are related by
\begin{equation}
\begin{aligned}
\label{eq:rmetricNHtrans}
    \delta G_{rr}&\,=e^{i\omega(t-v)}\left(\delta\mathcal{G}_{rr}+\frac{2\delta\mathcal{G}_{rv}}{r^2f}+\frac{\delta\mathcal{G}_{vv}}{r^4f^2}\right),\quad\quad
    \delta G_{rt}=e^{i\omega(t-v)}\left(\delta\mathcal{G}_{rv}+\frac{\delta\mathcal{G}_{vv}}{r^2f}\right),\\
    \delta G_{ri}&\,=e^{i\omega(t-v)}\left(\delta\mathcal{G}_{ri}+\frac{\delta\mathcal{G}_{vi}}{r^2f}\right).
    \end{aligned}
\end{equation}
With our decomposition \eqref{2.8} of the radial components of the metric appearing on the left hand side, regularity of the components in ingoing coordinates fixes the near-horizon gauge transformations $\xi_a$. These expressions are lengthy and uninformative so we will not write them out explicitly.

\paragraph{}By substituting the expressions for $\xi_a$ into the fourteen regularity conditions \eqref{eq:genericNHBC}, we obtain regularity conditions on the fields $\hat{g}_{\mu\nu}$ and $\hat{h}_{\mu\nu\rho}^I$ that we solved for above. Specifically we find that in the transverse sector the near-horizon expansions of the gauge-invariant fields $\partial_t\hat{g}^x_y-\partial_x\hat{g}^y_t$ and $\hat{h}_{txy}^1$ must satisfy \eqref{eq:genericNHBC}:
\begin{equation}
\label{eq:transverseregularityconds}
    \Gamma[\partial_t\hat{g}^x_y-\partial_x\hat{g}^y_t]=-\frac{i\omega}{4\pi T}\alpha[\partial_t\hat{g}^x_y-\partial_x\hat{g}^y_t],\quad\quad\quad\quad \Gamma[\hat{h}_{txy}^1]=-\frac{i\omega}{4\pi T}\alpha[\hat{h}_{txy}^1].
\end{equation}
In the longitudinal sector, the near-horizon expansions of the gauge-invariant fields $\partial_t \hat{g}_- +\partial_x\hat{g}^x_t$ and $\hat{h}_{txy}^2$ must satisfy \eqref{eq:genericNHBC} also:
\begin{equation}
\label{eq:longitudinalregularityconds}
        \Gamma[\partial_t\hat{g}_- +\partial_x\hat{g}^x_t]=-\frac{i\omega}{4\pi T}\alpha[\partial_t\hat{g}_- +\partial_x\hat{g}^x_t],\quad\quad\quad\quad \Gamma[\hat{h}_{txy}^2]=-\frac{i\omega}{4\pi T}\alpha[\hat{h}_{txy}^2].
\end{equation}
These are the only four relations imposed by the conditions \eqref{eq:genericNHBC}: the other conditions are automatically satisfied either identically or on-shell.

\subsubsection{Transport coefficients}

\paragraph{}The boundary conditions \eqref{eq:transverseregularityconds} and \eqref{eq:longitudinalregularityconds} provide four additional relations between the field theory sources and expectation values, beyond the conservation equations described above. These relations, combined with the Ward identity \eqref{eq:traceWardidentityperturbed}, give the hydrodynamic constitutive relations.

\paragraph{}More precisely, the five relations \eqref{eq:transverseregularityconds}, \eqref{eq:longitudinalregularityconds} and \eqref{eq:traceWardidentityperturbed} can be solved to give expressions for the currents $\langle\delta T_\pm\rangle$, $\langle\delta T^{x}_{\;\; y}\rangle$ and $\langle\delta J_I^{xy}\rangle$ in terms of the conserved charge densities $\langle \delta T^t_{\;\;\mu}\rangle$, $\langle\delta J_I^{t\mu}\rangle$ and sources in a derivative expansion. Substituting in the constitutive relations \eqref{eq:transversechargeconsrels} and \eqref{eq:longitudinalchargeconsrels} for the charge densities in terms of the hydrodynamic variables into these expressions yields the (time derivative of) the viscoelastic constitutive relations \eqref{eq:transversecurrentconsrels} and \eqref{eq:longitudinalcurrentconsrels} for the currents with the following values of the transport coefficients
\begin{equation}   
\begin{aligned}
\label{eq:transportcoeffsholoresults}
G=m^2(\mathcal{M}+\theta),\quad\quad\quad \eta=\frac{s}{4\pi}\Theta_1(r_h)^2,\quad\quad\quad \zeta=0,\quad\quad\quad \sigma=\frac{4\pi sT^2}{\chi^2}.
\end{aligned}
\end{equation}
It is in fact the direct relations between the currents and charge densities \eqref{eq:transverseregularityconds}, \eqref{eq:longitudinalregularityconds} and \eqref{eq:traceWardidentityperturbed} that are physical and define the values of the transport coefficients -- the constitutive relations presented before rely on a specific choice of hydrodynamic frame and this is why we had to put in \eqref{eq:transversechargeconsrels} and \eqref{eq:longitudinalchargeconsrels} by hand to reproduce them.

\paragraph{}Expressions of the form \eqref{eq:transportcoeffsholoresults} have been derived previously in other holographic models of viscoelasticity starting from Kubo formulae (see, for example, 
\cite{Amoretti:2019cef}). The important difference (besides the different holographic theory) is that we are not starting from Kubo formulae, which rely on assuming a certain theory of hydrodynamics. Instead we have derived the hydrodynamic theory (and hence also the Kubo formulae) directly from gravity.

\paragraph{}The vanishing of the bulk viscosity is not due to conformal symmetry, which is explicitly broken by $\mathcal{M}$. However, the simple way in which conformal symmetry is broken -- by a double-trace deformation that enters in the gravitational description as a modification of the boundary conditions --  is presumably the reason that $\zeta$ still vanishes. The three non-vanishing transport coefficients in \eqref{eq:transportcoeffsholoresults} depend explicitly on the double-trace coupling $\mathcal{M}$.

\paragraph{}The first order transport coefficients $\eta$, $\zeta$ and $\sigma$ are all manifestly non-negative, as is required for stability of the equilibrium state. However, this is not the case for the shear modulus $G$ and we will see later that $G$ is indeed negative in regions of parameter space. Note also that the shear viscosity $\eta$ is generically not equal to $s/4\pi$. In fact, we will see later that it is always less than this ratio. 

\subsection{The shear modulus and shear viscosity}
\label{sec:Thetasols}

\paragraph{}We have shown that the dynamics of small amplitude and long wavelength perturbations of the black hole are governed by the theory of viscoelastic hydrodynamics with specific values of the transport coefficients. We have given explicit expressions for all of these transport coefficients besides the shear modulus $G$ and the shear viscosity $\eta$. These coefficients are governed by solutions to the differential equation
\begin{equation}
\label{eq:Theta1Eq}
    \pa_r \left( r^4f \pa_r \Theta_1 \right) - \nn^2 \Theta_1 = 0,
\end{equation}
See equation \eqref{eq:transportcoeffsholoresults} and the discussion around equation \eqref{3.25} above. In this Section we are going to give explicit expressions for these coefficients in the limits of low temperature, low density as well as at the `self-dual' point $m^2=2r_h^2$. We will then solve the equation numerically to determine these transport coefficients over the full parameter space.

\subsubsection{Low temperature limit}

\paragraph{}Recalling the expression \eqref{5.4b} for the equilibrium temperature, we can access the low temperature limit by solving the equation \eqref{eq:Theta1Eq} perturbatively in powers of $\left(6-\frac{m^2}{r_h^2}\right)$:
\begin{equation}
    \Theta_1(r)=\Theta_1^{(0)}(r)+\left(6-\frac{m^2}{r_h^2}\right)\Theta_1^{(1)}(r)+\ldots.
\end{equation}
We impose that the solution at each order is finite at the horizon and that $\Theta_1^{(0)}(\infty)=1$ with $\Theta_1^{(n)}(\infty)=0$ for $n>0$. The solutions at the first two orders are
\begin{equation}
    \begin{aligned}
    \Theta_1^{(0)}&\,=\frac{2(2r-5r_h)(r-r_h)+3(r^2-2rr_h-2r_h^2)\log\left(\frac{r+2r_h}{3r}\right)}{(r-r_h)^2(4-3\log3)},\\
    \Theta_1^{(1)}&\,=\frac{r_h^2\left(2r^2-rr_h-r_h^2+\frac{3}{2}r(r+2r_h)\log\left(\frac{r+2r_h}{3r}\right)\right)}{(r-r_h)^3(r+2r_h)(4-3\log3)}.
    \end{aligned}
\end{equation}
From the near-boundary and near-horizon expansions of these we then obtain
\begin{equation}
    \begin{aligned}
    \label{eq:highdensitythetaTheta1}
        \Theta_1(r_h)&\,=\frac{2(6r_h^2-m^2)}{27r_h^2(4-3\log3)}+\ldots,\quad\quad\quad\quad 
        \theta=\frac{3r_h(2r_h^2-m^2)}{2m^2}\frac{9\log3-8}{4-3\log3}+\ldots,        
    \end{aligned}
\end{equation}
which gives the following low temperature shear modulus and shear viscosity
\begin{equation}
    \begin{aligned}
    \label{eq:ShearModulusLargeDensity}
        G&\,=m^2\left(\mathcal{M}-\frac{m(9\log3-8)}{\sqrt{6}(4-3\log3)}\right)+\ldots,\quad
        \eta=\left(\frac{16\pi }{27(4-3\log3)}\right)^2T^2+\ldots.
    \end{aligned}
\end{equation}

\subsubsection{Low density limit}

\paragraph{}To access the low density limit we expand the field perturbatively in powers of $m^2$ keeping $r_h$ fixed:
\begin{equation}
    \Theta_1(r)=\Theta_1^{(0)}(r)+m^2\Theta_1^{(1)}(r)+m^4\Theta_1^{(2)}(r)+\ldots,
\end{equation}
and then solve \eqref{eq:Theta1Eq} order-by-order in this expansion, imposing that the solution at each order is finite at the horizon and that $\Theta_1^{(0)}(\infty)=1$ with $\Theta_1^{(n)}(\infty)=0$ for $n>0$. The leading order solution is just $\Theta_1^{(0)}(r)=1$ and the first correction is
\begin{equation}
    \Theta_1^{(1)}(r)=\frac{1}{2\sqrt{3}r_h^2}\left(\pi-2\tan^{-1}\left(\frac{2r+r_h}{\sqrt{3}r_h}\right)-\sqrt{3}\log\left(\frac{r^2+rr_h+r_h^2}{r^2}\right)\right).
\end{equation}
From this we can obtain the first two terms in the small density expansion of $\Theta_1(r_h)$
\begin{equation}
\label{eq:lowdensityTheta1}
\Theta_1(r_h)=1+\frac{m^2}{6r_h^2}\left(\frac{\pi}{\sqrt{3}}-3\log3\right)+O(m^4),
\end{equation}
and therefore of the shear viscosity
\begin{equation}
\begin{aligned}
    \eta&\,=\frac{s}{4\pi}+\frac{m^2}{3}\left(\frac{\pi}{\sqrt{3}}-3\log3\right)+\ldots.
\end{aligned}    
\end{equation}
To obtain the first two terms in the expansion of the shear modulus, we need to work to one higher order in the expansion of $\Theta_1(r)$. In fact, it will be sufficient to determine the derivative of this correction, which is
\begin{equation}
\begin{aligned}
    \frac{d\Theta_1^{(2)}}{dr}=\frac{1}{2r_h^2(r^3-r_h^3)}\Biggl(&\,\frac{r_h}{r}\log3+\frac{\pi\left(1+\frac{r_h}{r}\right)}{\sqrt{3}}+\frac{r_h^2(r-r_h)}{r(r^2+rr_h+r_h^2)}\\
    &\,-\frac{2\left(1+\frac{2r_h}{r}\right)}{\sqrt{3}}\tan^{-1}\left(\frac{2r+r_h}{\sqrt{3}r_h}\right)-\log\left(1+\frac{r_h}{r}+\frac{r_h^2}{r^2}\right)\Biggr).
\end{aligned}
\end{equation}
Expanding near the boundary yields
\begin{equation}
\label{eq:lowdensitytheta}
    \theta=-r_h-\frac{m^2}{2r_h}\left(\frac{\pi}{\sqrt{3}}-\log3\right)+O(m^4),
\end{equation}
and therefore
\begin{equation}
\label{eq:ShearModulusLowDensity}
    G=m^2\left(\mathcal{M}-r_h-\frac{m^2}{2r_h}\left(\frac{\pi}{\sqrt{3}}-\log3\right)\right)+\ldots.
\end{equation}

\subsubsection{Self-dual point}

\paragraph{}The final limit in which we can analytically determine the shear modulus and shear viscosity is when $m^2=2r_h^2$ (or equivalently $m^2=8\pi^2T^2$). For this value there is an emergent self-duality in the bulk perturbation equations, as well as a simplification in the emblackening function such that these equations can be solved explicitly in terms of hypergeometric functions \cite{Davison:2014lua}. 

\paragraph{}Specifically, in this case
\begin{equation}
\begin{aligned}
    \Theta_1(r)=&\,\frac{4i}{\sqrt{7\pi}}\Biggl(\frac{\Gamma\left(\frac{5+\sqrt{7}i}{4}\right)\Gamma\left(1-\frac{\sqrt{7}i}{2}\right)}{\Gamma\left(\frac{-1-\sqrt{7}i}{4}\right)}\left(\frac{r}{r_h}\right)^{-\frac{1}{2}+\frac{\sqrt{7}i}{2}
    }{}_{2}F_{1}\left(-\frac{1}{4}+\frac{\sqrt{7}i}{4},\frac{5}{4}+\frac{\sqrt{7}i}{4},1+\frac{\sqrt{7}i}{2},\frac{r^2}{r_h^2}\right)\\
    &\,-\frac{\Gamma\left(\frac{5-\sqrt{7}i}{4}\right)\Gamma\left(1+\frac{\sqrt{7}i}{2}\right)}{\Gamma\left(\frac{-1+\sqrt{7}i}{4}\right)}\left(\frac{r}{r_h}\right)^{-\frac{1}{2}-\frac{\sqrt{7}i}{2}
    }{}_{2}F_{1}\left(-\frac{1}{4}-\frac{\sqrt{7}i}{4},\frac{5}{4}-\frac{\sqrt{7}i}{4},1-\frac{\sqrt{7}i}{2},\frac{r^2}{r_h^2}\right)\Biggr),
    \end{aligned}
\end{equation}
is the solution satisfying $\Theta_1(\infty)=1$. Expanding it near the horizon and boundary gives
\begin{equation}
\label{eq:thetaselfdual}
    \Theta_1(r_h)=\frac{4\pi^{3/2}\sech\left(\frac{\sqrt{7}\pi}{2}\right)}{\left|\Gamma\left(-\frac{1}{4}+\frac{\sqrt{7}i}{4}\right)\right|^2},\quad\quad\quad\quad\quad\quad \theta=-\frac{8\pi^2r_h\sech\left(\frac{\sqrt{7}\pi}{2}\right)}{\left|\Gamma\left(-\frac{1}{4}+\frac{\sqrt{7}i}{4}\right)\right|^4}.
\end{equation}
These give the following shear modulus and shear viscosity
\begin{equation}
    G=2(2\pi T)^2\left(\mathcal{M}- T\frac{16\pi^3\sech\left(\frac{\sqrt{7}\pi}{2}\right)}{\left|\Gamma\left(-\frac{1}{4}+\frac{\sqrt{7}i}{4}\right)\right|^4}\right),\quad\quad\quad\quad \eta=\frac{16\pi^{3}\sech^2\left(\frac{\sqrt{7}\pi}{2}\right)}{\left|\Gamma\left(-\frac{1}{4}+\frac{\sqrt{7}i}{4}\right)\right|^4}(2\pi T)^2.
\end{equation}

\paragraph{}In the context of viscoelastic hydrodynamics, the case $m^2=8\pi^2T^2$ is special in that the speeds of the transverse and longitudinal modes are equal at this point
\begin{equation}
    v_\perp^2=v_\parallel^2=\frac{G}{2\mathcal{M}r_h^2}=1-\frac{T}{\mathcal{M}}\frac{16\pi^3\sech\left(\frac{\sqrt{7}\pi}{2}\right)}{\left|\Gamma\left(-\frac{1}{4}+\frac{\sqrt{7}i}{4}\right)\right|^4}.
\end{equation}

\subsubsection{Numerical solution}

\begin{figure}[H]
	\centering
	\begin{subfigure}{.5\textwidth}
		\centering
		\includegraphics[width=1.0\linewidth]{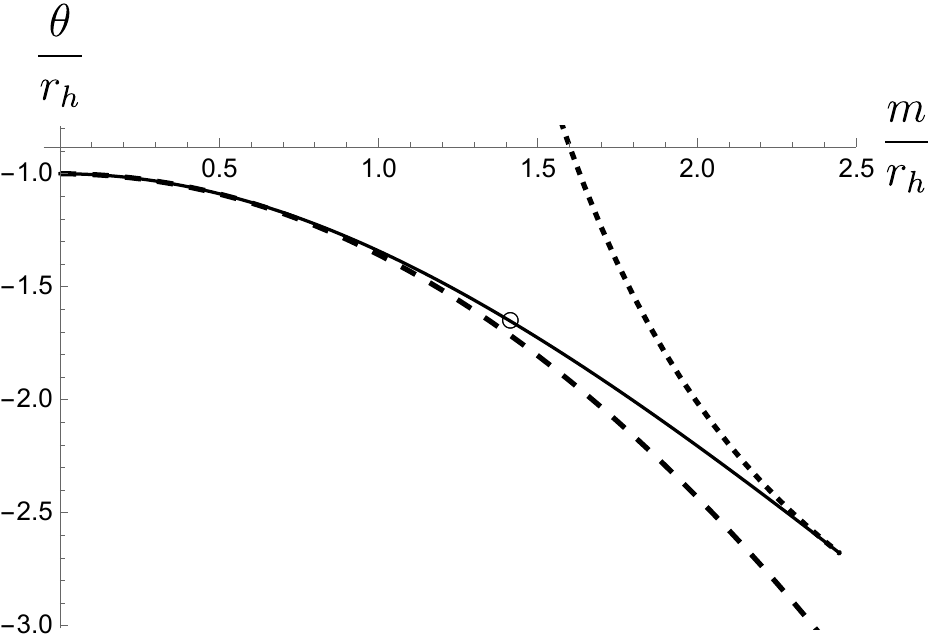}
	\end{subfigure}%
	\begin{subfigure}{.5\textwidth}
		\centering
		\includegraphics[width=1.0\linewidth]{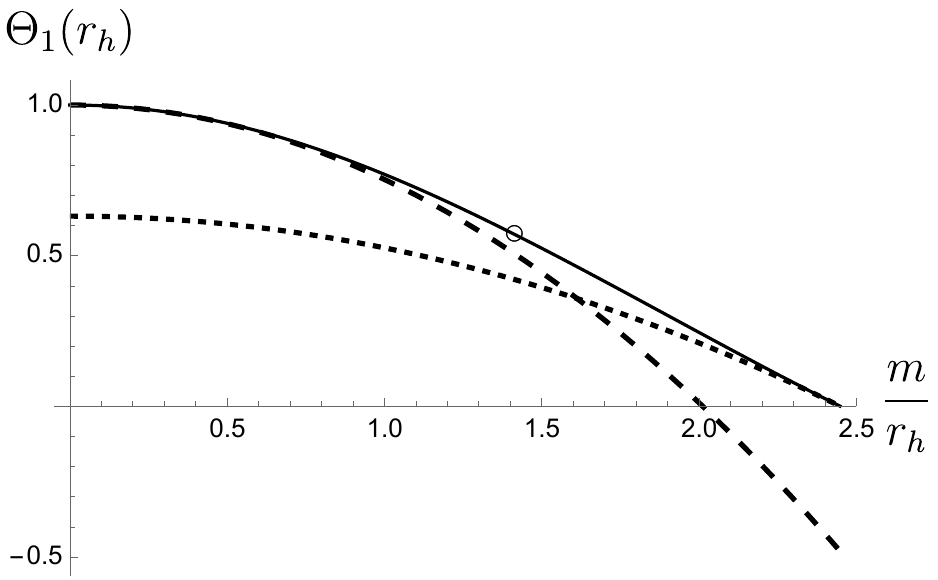}
	\end{subfigure}
	\caption{Plots of $\theta/r_h$ (left) and $\Theta_1(r_h)$ (right) as a function of $m/r_h$. Note that the right hand edge of each plot $m=\sqrt{6}r_h$ corresponds to the limit $m/T\rightarrow\infty$. The solid lines are numerical solutions of \eqref{eq:Theta1Eq} while the dashed lines are the analytic approximations at low density (equations \eqref{eq:lowdensitytheta} and \eqref{eq:lowdensityTheta1}) and low temperature (equation \eqref{eq:highdensitythetaTheta1}). The circle is the result \eqref{eq:thetaselfdual} for $m=\sqrt{2}r_h$.}
\label{fig:Theta1SolutionsNumerical}
\end{figure}

\paragraph{}Beyond the limits above where analytic expressions are accessible, the equation \eqref{eq:Theta1Eq} can easily be solved numerically to obtain the transport coefficients for any temperature and density. In Figure \ref{fig:Theta1SolutionsNumerical} below we show numerical results for the quantities $\theta$ and $\Theta_1(r_h)$ that control the shear modulus and viscosity. They agree very well with the analytic approximations in the appropriate limits, and interpolate straightforwardly between these limits. 

\paragraph{}These plots exhibit two important features. Firstly, notice that $\Theta_1(r_h)\leq1$ over the full parameter space. This can also be shown analytically using a version of the argument in Section 3 of \cite{Hartnoll:2016tri}. It means that the shear viscosity $\eta\leq\frac{s}{4\pi}$ \cite{Kovtun:2004de} -- in contrast to the usual holographic expectations -- with equality only in the zero density limit. This is similar to other holographic models governed by viscoelastic hydrodynamics \cite{Alberte:2016xja,Alberte:2017oqx,Andrade:2019zey,Amoretti:2019cef,Li:2024rzd}.

\paragraph{}Secondly, notice that $\theta$ is negative over the full parameter space and its magnitude grows monotonically as $m/T$ is increased. This is crucial: a negative $\theta$ gives a negative contribution to the shear modulus $G$. From equation \eqref{eq:transversedispersion} a negative $G$ corresponds to a negative squared speed of the transverse sound mode $v_\perp^2$, signalling an instability of the equilibrium state. We will explore this, and other instabilities, in detail in the following Section.

\section{Hydrodynamic modes and instabilities}
\label{sec:instabilities}

\paragraph{}We have established that the small amplitude, long wavelength perturbations of the black brane are governed by the viscoelastic hydrodynamics of Section \ref{sec:HydroSec}. In particular, this guarantees that the black hole quasinormal modes in this regime are simply the hydrodynamic modes \eqref{eq:transversedispers}, \eqref{eq:sounddispers} and \eqref{eq:diffusiondispers}, with the values of the transport coefficients calculated above. It is instructive to examine these hydrodynamic modes more carefully: stability of the equilibrium state to small amplitude perturbations requires that the transport coefficients $v_{\perp,\parallel}^2$, $\Gamma_{\perp,\parallel}$ and $D_\parallel$ all be non-negative. We will find that this is violated over a large part of the parameter space, implying perturbative instability of the equilibrium black hole.

\paragraph{}The transport coefficients, and hence the hydrodynamic modes, depend on three dimensionful parameters: $m$, $T$ and $\mathcal{M}$. It will be useful to organise the analysis by considering how quantities depend on $m/T$ for a fixed $\mathcal{M}/r_h$. Recalling the expression \eqref{5.4b} for the temperature, this can be achieved by studying the behaviour of quantities as a function of $0<m/r_h<\sqrt{6}$ for fixed $\mathcal{M}/r_h$. 

\subsection{Transverse modes}

\begin{figure}[H]
	\centering
	\begin{subfigure}{.5\textwidth}
		\centering
		\includegraphics[width=1.0\linewidth]{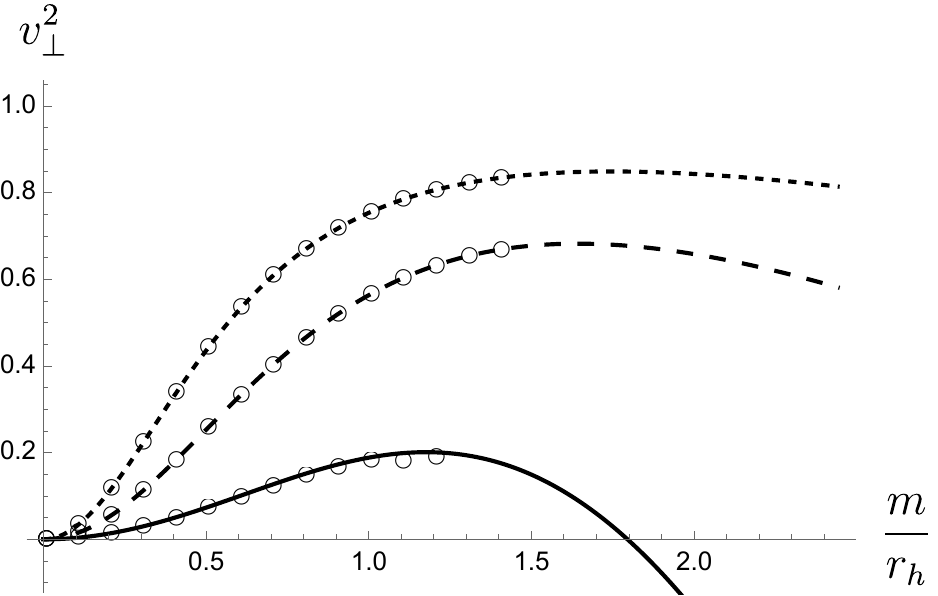}
	\end{subfigure}%
	\begin{subfigure}{.5\textwidth}
		\centering
        \includegraphics[width=1.0\linewidth]{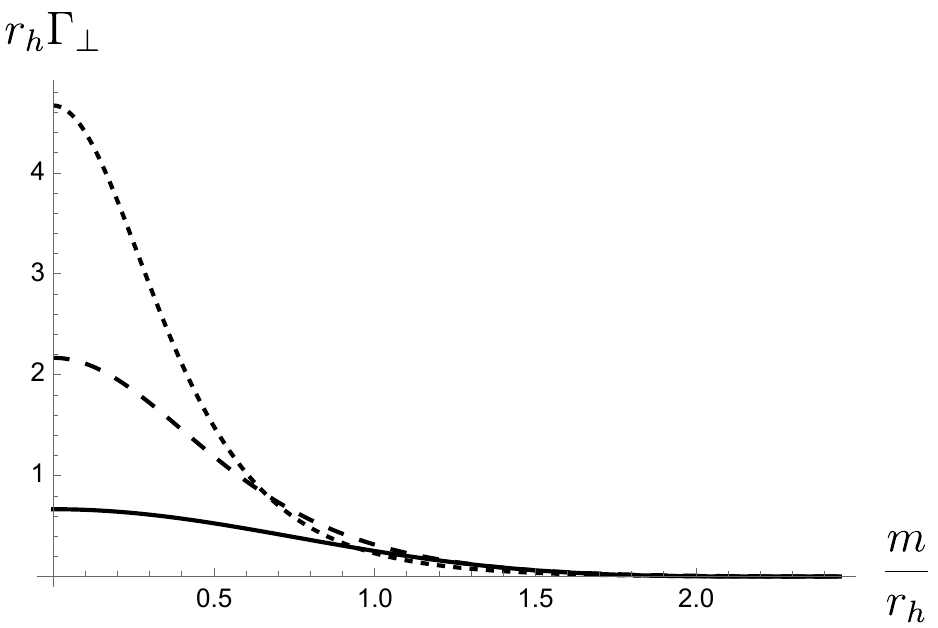}
	\end{subfigure}
	\caption{The squared speed $v_\perp^2$ and attenuation $\Gamma_\perp$ of the hydrodynamic modes obtained using the transport coefficients calculated in Section \ref{sec:holography}. Results are shown for $\mathcal{M}/r_h=10$ (short-dashed), $\mathcal{M}/r_h=5$ (long-dashed) and $\mathcal{M}/r_h=2$ (solid). The circles are the results of \cite{Grozdanov:2018ewh} from explicit numerical computation of black hole quasinormal modes.}
\label{fig:TransverseModes1}
\end{figure}

\paragraph{}We start with the transverse sound modes which have dispersion relations \eqref{eq:transversedispersion}. The squared speed of these modes is related to the transport coefficients by $v_\perp^2=G/\chi$. In the left panel of Figure~\ref{fig:TransverseModes1} we plot the speed of these modes using the transport coefficients calculated in Section \ref{sec:holography}. We compare these results to the explicit numerical computations of the quasinormal modes of this black hole in \cite{Grozdanov:2018ewh}, finding excellent agreement. In particular, as anticipated in \cite{Armas:2019sbe}, using the higher-form hydrodynamics of Section \ref{sec:HydroSec} resolves the discrepancies visible in \cite{Grozdanov:2018ewh} between quasinormal mode calculations and a more primitive theory of hydrodynamics.

\paragraph{}The computation of quasinormal modes in \cite{Grozdanov:2018ewh} also highlighted an instability of the equilibrium state at low density. Specifically, that for $\mathcal{M}<r_h$ (which will always be the case at sufficiently high temperature or density) the squared speed of the transverse sound mode $v_\perp^2<0$. We can understand this instability more fundamentally by examining the shear modulus $G$, which in the language of field theory is inversely proportional to the static susceptibility of the one-form charge density
\begin{equation}
    G^{-1}=\frac{1}{Q^2\gamma^2}\lim_{\omega\rightarrow0}\langle J_1^{ty}J_1^{ty}\rangle.
\end{equation}
This is required to be non-negative for a stable equilibrium state. The low density expansion of the shear modulus is given in \eqref{eq:ShearModulusLowDensity}, where the leading term is seen to be non-negative only for $\mathcal{M}>r_h$.

\paragraph{}We will now explore what happens beyond the low density limit. In particular we will identify the range of values of $\mathcal{M}/r_h$ for which the equilibrium state is stable for all densities $m/T$. First, notice that the momentum susceptibility
\begin{equation}
    \chi=sT+m^2(\mathcal{M}-r_h),
\end{equation}
is always positive for $\mathcal{M}>r_h$ and so a negative $v_\perp^2$ must come from a negative shear modulus $G$. The first correction to $G$ in the small density expansion \eqref{eq:ShearModulusLowDensity} is negative, indicating that the condition $\mathcal{M}>r_h$ is not sufficient to ensure stability at higher density. This is confirmed by the earlier numerical results for $\theta$ in Figure~\ref{fig:Theta1SolutionsNumerical}: the magnitude of its (negative) contribution to $G$ increases monotonically as the density increases. In other words, the most restrictive condition on $\mathcal{M}/r_h$ to ensure stability at all densities comes from demanding that $G$ be non-negative at large density $m/T\rightarrow\infty$. Using the large density expansion for the shear modulus \eqref{eq:ShearModulusLargeDensity}, this is the case provided that
\begin{equation}
\label{eq:transversestabilitythreshold}
    \frac{\mathcal{M}}{r_h}\geq\frac{(9\log3-8)}{(4-3\log3)}\approx 2.68.
\end{equation}
This is consistent with the results shown in the left panel of Figure~\ref{fig:TransverseModes1}.

\paragraph{}The transverse sound modes must also be subluminal. Using our holographic expressions for the transport coefficients, this requires that
\begin{equation}
\label{eq:causalitycond1}
\frac{\theta}{r_h}\leq3\left(\frac{r_h^2}{m^2}-\frac{1}{2}\right).
\end{equation}
This condition is independent of $\mathcal{M}$. Using the earlier results for $\theta$ in Figure~\ref{fig:Theta1SolutionsNumerical} we have verified that this is satisfied for all $m/T$.

\paragraph{}Stability also requires that the attenuation of the transverse sound mode $\Gamma_\perp$ in be non-negative. Substituting the holographic expression \eqref{eq:transportcoeffsholoresults} for $\sigma$ into \eqref{eq:transversedispersion}, this can be written as
\begin{equation}
    \Gamma_\perp=\frac{2\pi sT^2}{m^2\chi^2}\left(\frac{m^2\chi}{4\pi sT^2}\eta+G\right).
\end{equation}
Since the first term in brackets is always positive in the stable region identified above, there is no way for $\Gamma_\perp$ to be negative unless $G$ (and hence $v_\perp^2$) already is and so the threshold for stability of the transverse modes is \eqref{eq:transversestabilitythreshold}. In the right panel of Figure~\ref{fig:TransverseModes1} we show $\Gamma_\perp$ computed using the transport coefficients of Section \ref{sec:holography}.

\subsection{Longitudinal modes}

\begin{figure}[H] 
\centering
\begin{subfigure}{.5\textwidth}
	\centering
	\includegraphics[width=1.0\linewidth]{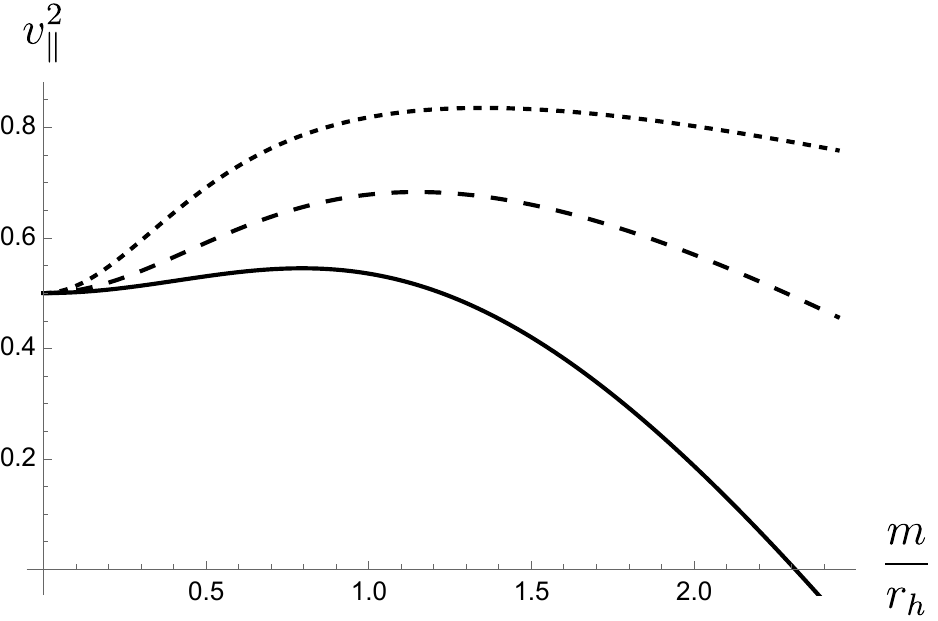}
\end{subfigure}%
\begin{subfigure}{.5\textwidth}
	\centering
\includegraphics[width=1.0\linewidth]{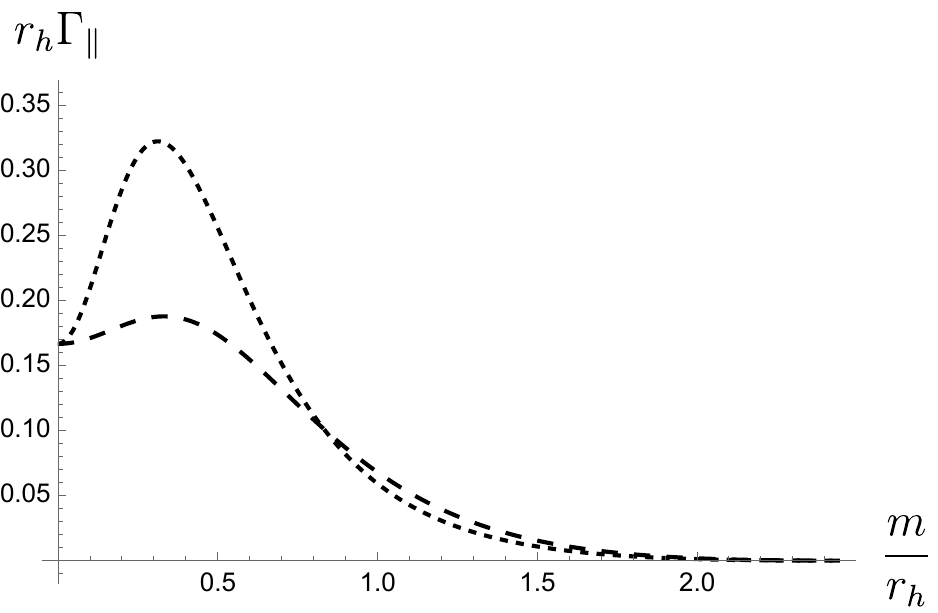}
\end{subfigure}
\caption{The squared speed $v_\parallel^2$ and attenuation $\Gamma_\parallel$ of the hydrodynamic modes obtained using the transport coefficients calculated in Section \ref{sec:holography}. Results are shown for $\mathcal{M}/r_h=10$ (short-dashed), $\mathcal{M}/r_h=5$ (long-dashed) and $\mathcal{M}/r_h=3$ (solid). The attenuation is not shown for the latter case: it has a pole where the speed vanishes.}
\label{fig:LongitudinalModes1}
\end{figure}

\paragraph{}In addition to the two transverse sound modes, there are longitudinal sound modes with dispersion relations \eqref{eq:sounddispers}. Using the expressions \eqref{eq:transportcoeffsholoresults} for the transport coefficients for the holographic theory, the speed and attenuation of these modes can be written
\begin{equation}
\label{eq:stabilitylongsound}
    v_\parallel^2=v_\perp^2-\frac{3r_h(m^2-2r_h^2)}{4\chi},\quad\quad\quad\quad\Gamma_\parallel=\frac{1}{2}\left(\frac{\eta}{\chi}+\frac{4\pi sT^2}{m^2\chi}\frac{(1-2v_\parallel^2)^2}{4v_\parallel^2}\right).
\end{equation}
In the left panel of Figure~\ref{fig:LongitudinalModes1} we show results for $v_\parallel^2$ as a function of density. 

\paragraph{}Just like for the transverse sound modes, there are instabilities of the equilibrium state that are reflected in negative values of $v_\parallel^2$. In fact, the longitudinal sound modes are unstable even when the transverse ones are not, and so these give a more restrictive condition for stability of the state. Schematically this is because at high densities -- where $v_\perp^2$ is smallest -- we see from equation \eqref{eq:stabilitylongsound} that $v_\parallel^2<v_\perp^2$. More precisely, the condition $v_\parallel^2>0$ corresponds to
\begin{equation}
\label{eq:longstabcond1}
    \frac{\mathcal{M}+\theta}{r_h}-\frac{3}{4}\left(1-\frac{2r_h^2}{m^2}\right)>0.
\end{equation}
Recalling that $\theta/r_h$ decreases monotonically as density is increased, for a fixed $\mathcal{M}/r_h$ both terms on the left hand side are smallest at high density. Therefore to find the threshold value of $\mathcal{M}/r_h$ for which the state is stable for all $m/T$, we need to look at the high density limit. Using the high density expansion \eqref{eq:highdensitythetaTheta1}, the stability condition \eqref{eq:longstabcond1} is
\begin{equation}
\label{eq:longitudinalstabilitythreshold}
    \frac{\mathcal{M}}{r_h}>\left(\frac{1}{2}+\frac{(9\log3-8)}{(4-3\log3)}\right)\approx 3.18,
\end{equation}
which is more restrictive than \eqref{eq:transversestabilitythreshold} found before. In the left panel of Figure~\ref{fig:LongitudinalModes1} we show $v_\parallel^2$ for values of $\mathcal{M}$ near the threshold \eqref{eq:longitudinalstabilitythreshold}.

\paragraph{}The condition that the longitudinal sound modes are subluminal may be written as 
\begin{equation}
\frac{\theta}{r_h}\leq\frac{3}{2}\left(\frac{r_h^2}{m^2}-\frac{1}{2}\right).
\end{equation}
Again, this is independent of $\mathcal{M}$. Recalling that $\theta$ is negative, this is satisfied for all $m/T$ since \eqref{eq:causalitycond1} is.

\paragraph{}In the right panel of Figure~\ref{fig:LongitudinalModes1} we show the attenuation $\Gamma_\parallel$ computed using the transport coefficients of the holographic theory. It is manifest from equation \eqref{eq:stabilitylongsound} that this is always positive in the stable region $v_\parallel^2>0$ and so is not the source of any additional instability.

\begin{figure}[H]
\centering
\includegraphics[width=0.5\linewidth]{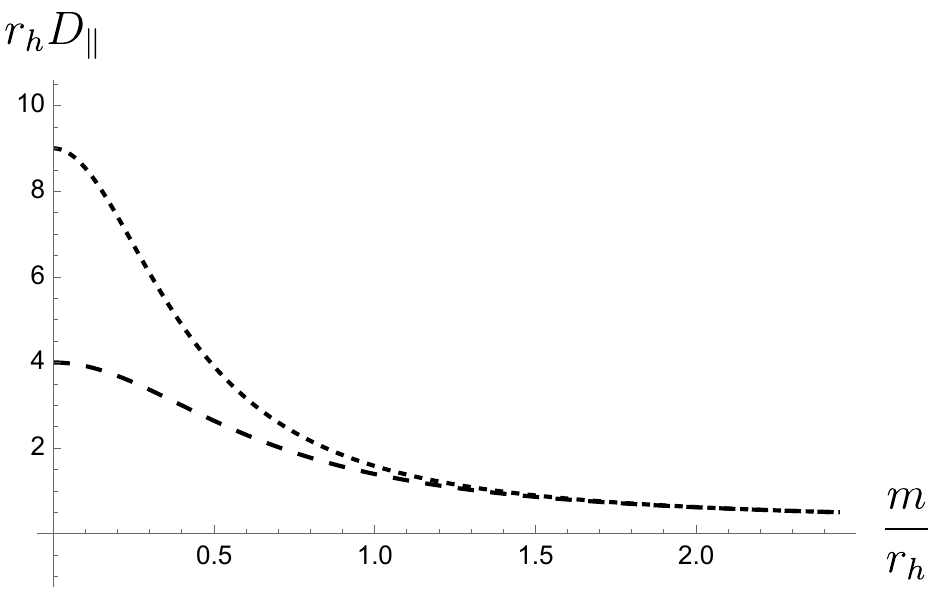}
\caption{The hydrodynamic diffusivity $D_\parallel$ obtained using the transport coefficients calculated in Section \ref{sec:holography}. Results are shown for $\mathcal{M}/r_h=10$ (short-dashed), $\mathcal{M}/r_h=5$ (long-dashed).}
\label{fig:LongitudinalModes2}
\end{figure}

\paragraph{}The final hydrodynamic mode is the diffusive mode with dispersion relation \eqref{eq:diffusiondispers}. Using our expressions \eqref{eq:transportcoeffsholoresults} for the transport coefficients of the holographic theory, the diffusivity is
\begin{equation}
    D_\parallel= \frac{4\pi sT}{m^2\chi c_\rho v_\parallel^2}(B + G-Q\gamma^2),
\end{equation}
and in Figure~\ref{fig:LongitudinalModes2} we show $D_\parallel$ using the transport coefficients computed in Section \ref{sec:holography}. 
For stability of the equilibrium state we require $D_\parallel>0$. When $v_\parallel^2>0$, the only way this can be violated is by a sufficiently negative shear modulus $G$. More precisely, the stability condition is that
\begin{equation}
\label{eq:stabilitycond3}
    \frac{\mathcal{M}+\theta}{r_h} -\frac{m^2}{6r_h^2+m^2}>0.
\end{equation}
Recalling that $\theta/r_h$ decreases monotonically as density is increased, for a fixed $\mathcal{M}/r_h$ both terms on the left hand side are smallest at high density. In fact, at high density the inequality is equivalent to that in \eqref{eq:longstabcond1} and so the stability condition is again \eqref{eq:longitudinalstabilitythreshold}. In summary, in order for all hydrodynamic modes to be stable for all $m/T$, the coupling $\mathcal{M}$ must satisfy the condition \eqref{eq:longitudinalstabilitythreshold}. Recalling the relation \eqref{eq:rhdefn} for $r_h(T,m)$, we see that at sufficiently low densities and temperatures, this condition will always be satisfied provided $\mathcal{M}\ne0$.

\paragraph{}Suppose we are in a stable equilibrium state: this requires the three inequalities \eqref{eq:longstabcond1}, \eqref{eq:stabilitycond3} and $G>0$ to be satisfied. It first becomes unstable when a small change in $T/\mathcal{M}$ or $m/\mathcal{M}$ causes one of the quantities on the left hand side of these inequalities to become negative. From their explicit form, it is straightforward to check that the quantity in the diffusivity inequality \eqref{eq:stabilitycond3} is less than or equal to the other two for any $m$, $T$ and $\mathcal{M}$. This means that the first instability to onset will always be that due to a negative diffusivity $D_\parallel$ of the incoherent charge density.

\paragraph{}From the Einstein relation, $D_\parallel$ is the ratio of the conductivity of the incoherent charge density \eqref{eq:incoherentdensitydef} that diffuses to its thermodynamic susceptibility. From viscoelastic hydrodynamics it is straightforward to obtain that the conductivity of this incoherent density is $\sigma$, which from \eqref{eq:transportcoeffsholoresults} is always positive in the holographic theory. Therefore the first instability to be triggered is fundamentally due to the susceptibility of the incoherent charge density \eqref{eq:incoherentdensitydef} becoming negative.

\section{Duality with diffusive hydrodynamics}
\label{sec:hodgeduality}

\paragraph{}Bulk solutions to the Einstein-Maxwell theory \eqref{lag} can be mapped to those of Einstein gravity coupled to massless scalar fields by using the Hodge duality operation. The scalar formulation of the solution \eqref{eq:BHmetric} has previously been studied, motivated by the physics of momentum relaxation \cite{Vegh:2013sk,Davison:2013jba,Andrade:2013gsa,Donos:2014cya,Davison:2015bea,Blake:2015epa}. Its hydrodynamics is significantly different than the higher-form hydrodynamics presented above. At long wavelengths there is a single hydrodynamic field and a single hydrodynamic mode: heat diffuses with diffusivity $D_T=\kappa/c$, with $\kappa$ the thermal conductivity and $c$ the specific heat. This difference ultimately stems from the different boundary conditions applied to the fields in each formulation. In this Section we will describe in detail how to obtain this diffusive hydrodynamics from the viscoelastic hydrodynamics of Section \ref{sec:HydroSec}. In particular, we will give explicit expressions for $\kappa$ and $c$ in terms of the viscoelastic transport coefficients. Furthermore, we will show that the heat diffusion mode of the scalar theory has a simple manifestation in viscoelastic hydrodynamics. It is the collective excitation of the viscoelastic fluid under no-flux boundary conditions \eqref{eq:nofluxdiffusivity}.

\subsection{Map between observables}

\paragraph{}The Hodge duality operation acts on the 3-form field strength as
\begin{equation}
\label{eq:hodgetransform}
    H^I_{bcd}\rightarrow\epsilon^{IJ}\sqrt{-G}\hat{\epsilon}_{abcd}F^a_J,
\end{equation}
and leaves the metric unchanged. After acting with this, the Einstein equations are the same as those obtained from the action
\begin{equation}
\label{eq:Einsteinscalarbulkaction}
    S_\Phi=\int d^4x\sqrt{-G}\left(R+6-\frac{1}{2}\sum_{I=1}^{2}F^I_aF^{Ia}\right),
\end{equation}
where $F_a^I$ are 1-form field strengths of scalar potentials $\Phi^I$
\begin{equation}
    F_a^I=\partial_a\Phi^I.
\end{equation}
This transformation also maps the Maxwell equations of one theory onto the Bianchi identities of the other.

\paragraph{}Applying this transformation to the equilibrium black hole solution of Section \ref{sec:holography}, the metric is as in \eqref{eq:BHmetric}, while the scalar field profiles are
\begin{equation}
\label{eq:scalarfieldeqmholo}
    \Phi^I=-mx^I.
\end{equation}
The scalar fields $\Phi^I$ are the bulk version of the crystal fields $\varphi^I$ of the theory of elasticity described in Section \ref{sec:HydroSec}. In particular, after imposing appropriate mixed boundary conditions on $\Phi_I$, their profiles \eqref{eq:scalarfieldeqmholo} correspond to the expectation values of scalar operators that spontaneously break translational symmetry. This has been explored in \cite{Armas:2019sbe} and we will not pursue it further here.

\paragraph{}Instead we want to understand the relation between our viscoelastic hydrodynamics and the bulk scalar theory \eqref{eq:Einsteinscalarbulkaction} with more conventional Dirichlet boundary conditions for the scalar fields. With these boundary conditions, the profiles \eqref{eq:scalarfieldeqmholo} correspond to sources for scalar operators in the field theory and so the translational symmetry is explicitly -- rather than spontaneously -- broken \cite{Andrade:2013gsa}. This distinction has a huge effect on the physics of the field theory. In particular, the only conserved charge is the total energy of the state and the theory of hydrodynamics governing this is simple: heat diffuses \cite{Davison:2014lua}.

\paragraph{}To understand how this diffusive hydrodynamics arises from higher-form hydrodynamics, we first need to understand how the Hodge transformation \eqref{eq:hodgetransform} relates the field theory sources and expectation values in the higher-form theory of Section \ref{sec:holography} to those in the scalar theory with Dirichlet boundary conditions. The holographic dictionary, summarised for each case in Appendix \ref{app:holodictionary}, relates these quantities to specific data in the near-boundary expansion of the bulk fields. Inserting these expansions into the map \eqref{eq:hodgetransform} then gives the following relations between the sources $b_{I\mu\nu}$ and expectation values $\langle J^{I\mu\nu}\rangle$ for the higher-form theory and the sources $\mathcal{J}_I$ and expectation values $\langle\mathcal{O}^I\rangle$ for the scalar theory with Dirichlet boundary conditions
\begin{equation}
\begin{aligned}
\label{eq:mattermap}
\braket{J_I^{\mu \nu}} & \rightarrow - \ep_{IJ} \varep^{\mu \nu \rho} \pa_\rho \mathcal{J}^J, \\
    (d b^I)_{\al \mu \nu} 	& \rightarrow \ep^{IJ} \varep_{\al \mu \nu} \braket{\mathcal{O}_J} - \ep^{IJ} \varep_{\al \mu \nu} \mathcal{M} \Box_g \mathcal{J}_J. 
\end{aligned}
\end{equation}
The mixing of sources and expectation values under this map  is the reason why the physics of the two theories are so different.

\paragraph{}The field theory metric in both cases is given by $g_{\mu\nu}$, but the energy-momentum tensor $\langle T^{\mu\nu}_\Phi\rangle$ in the scalar theory differs from that in the higher-form theory $\langle T^{\mu\nu}\rangle $:
\be \begin{aligned}
\label{eq:EMtensormap}
	\braket{T^{\mu \nu}} \rightarrow & \braket{T_\Phi^{\mu \nu}} - \mathcal{M} \left( \frac{g^{\mu \nu} g^{\rho \sig}}{2} - g^{\mu \rho} g^{\nu \sig} \right) \pa_\rho \mathcal{J}^I \pa_\sig \mathcal{J}^I.
\end{aligned} \ee
Using these, the non-vanishing equilibrium expectation values in the scalar theory are
\begin{equation}
\label{eq:scalarEQvalues}
    \langle T^{tt}_\Phi \rangle=-p+sT+2Q\gamma^2-\mathcal{M}m^2,\quad\quad\quad \langle T^{xx}_\Phi \rangle=\langle T^{yy}_\Phi \rangle=p-Q\gam^2,
\end{equation}
and the sources are $\mathcal{J}^I=-mx^I$.

\paragraph{}Although we are applying the map directly to the general theory of viscoelastic hydrodynamics of Section \ref{sec:HydroSec}, it is clearly valid only for the microscopic theories governed by the gravitational actions \eqref{lag} and \eqref{eq:Einsteinscalarbulkaction}. Therefore the expressions for expectation values in the scalar theory -- in equation \eqref{eq:scalarEQvalues} and below -- should be understood to be the particular functions of $T$, $m$ and $\mathcal{M}$ found by substituting in the specific viscoelastic thermodynamic and transport coefficients from Section \ref{sec:holography}. In fact, the double-trace coupling $\mathcal{M}$ does not exist in the scalar theory and so any quantity in this theory can only depend on combinations of viscoelastic transport coefficients in which $\mathcal{M}$ cancels out identically. This is the case for the right hand sides of \eqref{eq:scalarEQvalues} and, indeed, the explicit expressions in terms of $T$ and $m$ agree with those calculated directly in \cite{Andrade:2013gsa}.

\subsection{Diffusive hydrodynamics}

\paragraph{}The relations \eqref{eq:mattermap} and \eqref{eq:EMtensormap} tell us how to relabel the integration constants in the bulk solutions of Section \ref{sec:holography} so that they are expressed in terms of sources and expectation values of the QFT dual to the scalar theory with Dirichlet boundary conditions. From this, we can use our previous results for viscoelastic hydrodynamics to obtain diffusive hydrodynamics.

\paragraph{}Specifically, imposing the classical bulk equations for the higher-form theory requires that the energy-momentum tensor obeys the local conservation equation \eqref{2.3}. With the relabelling in equations \eqref{eq:mattermap} and \eqref{eq:EMtensormap}, this means that these equations in the scalar theory require that
\begin{equation} \label{eq:scalarEMconservation}
    	\na_\mu \langle T_\Phi^{\mu \nu}\rangle + \braket{\mathcal{O}_I} \pa^\nu \mathcal{J}^I=0,
\end{equation}
due to a cancellation between the $\mathcal{M}$-dependent terms. Using the same logic, the local conservation equation for the $1$-form charge \eqref{2.3} becomes a trivial identity upon the relabelling \eqref{eq:mattermap}. In this way, we have recovered the Ward identities governing the scalar formulation of the theory \cite{Andrade:2013gsa}. We now turn to obtaining the constitutive relations of this hydrodynamic theory.

\subsubsection{Regularity conditions}

\paragraph{}Unlike the Ward identities, the constitutive relations of hydrodynamics are not obtained solely from the general solutions to the bulk equations of motion. As we saw in Section \ref{sec:consrelationshigherform}, they instead arise after imposing ingoing boundary conditions on these solutions. We are therefore first going to check that, after applying the Hodge duality \eqref{eq:hodgetransform}, the ingoing solutions in the higher-form formulation become ingoing solutions in the scalar formulation. Having established this, we can then apply this map directly to the constitutive relations of viscoelastic hydrodynamics to obtain those of diffusive hydrodynamics.

\paragraph{}For this, we will specialise to the case of small amplitude perturbations and denote the perturbations of the one-form field strengths of the scalar potentials as $\delta F_a^I$. Under a local diffeomorphism $\xi^a$ this transforms as 
\begin{equation}
    \delta F_a^I\rightarrow\delta F_a^I +\xi^b \na_b F^I_{a} + F^I_{b} \na_{a} \xi^b ,    
\end{equation}
where $\nabla$ is the covariant derivative of the equilibrium metric \eqref{eq:BHmetric} and the field strength $F^I_a$ on the right hand side is that associated to the equilibrium potential \eqref{eq:scalarfieldeqmholo}. Analogously to the discussion around equation \eqref{2.8}, it is therefore convenient to parameterise this perturbation by
\begin{equation}
    \delta F_a^I=\hat{f}_a^I+m\del^I_b \partial_a\xi^b,
\end{equation}
where $d\hat{f}^I=0$. We perform a Fourier decomposition of the field perturbations analogous to that in \eqref{eq:fouriertransconvens}.

\paragraph{}We will now determine the ingoing boundary conditions for these fields. Upon transforming from the $t$ coordinate to the ingoing null coordinate $v$ defined in \eqref{eq:nullcoorddefn}, the Fourier components of the original field strength perturbations $\delta F^I_a$ are related to those in the ingoing coordinate system $\delta\mathcal{F}^I_a$ by
\begin{equation}
\delta F^I_{r} - \frac{\delta F^I_{t}}{r^2f} = e^{i \om (t-v)} \delta\mathcal{F}^I_{r}, \quad\quad\quad \delta F^I_{t} = e^{i \om (t-v)} \delta\mathcal{F}^I_{v}, \quad\quad\quad
\delta F^I_{i} = e^{i \om (t-v)} \delta\mathcal{F}^I_{i}, 
\end{equation}
where the $i$ index is for $\{x,y\}$. Ingoing boundary conditions at the horizon imposes the six conditions \eqref{eq:genericNHBC} on the objects on the left hand sides of these equations. Using the near-horizon gauge transformations $\xi_a$ described after \eqref{eq:rmetricNHtrans}, the conditions are satisfied provided
\be \begin{aligned} \label{10.20}
\Gamma[\hat{f}_I^{r}] & =-\frac{i\omega}{4\pi T}\alpha[\hat{f}_I^{r}],
\end{aligned} \ee
where $\hat{f}_I^{r}\equiv r^2f\hat{f}_r^{I}$ and $\Gamma$ and $\alpha$ denote the components of the near-horizon expansion of this field defined in equation \eqref{eq:genericNHexpn}. Once \eqref{10.20} is imposed, the remaining ingoing conditions are then automatically satisfied either identically or on-shell.

\paragraph{}The conditions \eqref{10.20} are equivalent to what one obtains by imposing the ingoing boundary conditions \eqref{eq:transverseregularityconds} and \eqref{eq:longitudinalregularityconds} for the higher-form fields $\hat{h}^I_{txy}$ and then performing the Hodge duality transformation \eqref{eq:hodgetransform}. To see this, we simply linearise the transformation \eqref{eq:hodgetransform} to obtain
\be \begin{aligned} \label{10.13}
\hat{h}^I_{txy} & \rightarrow r^2 \ep^{IJ}\hat{f}^r_J.
\end{aligned} \ee

\subsubsection{Constitutive relations}

\paragraph{}We have just established that the ingoing solutions for linearised perturbations of the scalar theory are simply the ingoing solutions to the higher-form theory after the relabelling \eqref{eq:mattermap} and \eqref{eq:EMtensormap}. We can therefore obtain the constitutive relations of diffusive hydrodynamics by applying this relabelling to the constitutive relations of viscoelastic hydrodynamics.

\paragraph{}It is instructive to derive diffusive hydrodynamics in a few steps. In the scalar theory there is no higher-form symmetry and so we are first going to integrate out the hydrodynamic fields associated to the higher-form currents, and their sources. To do this, we insert the viscoelastic hydrodynamic constitutive relations for $\langle\delta J^{\mu\nu}_I\rangle$ into the linearised left hand side of equation \eqref{eq:mattermap} and solve. For the transverse fields this gives
\begin{equation}
    \begin{aligned}
    \label{eq:transverseintegrateout}
    \delta \zeta_1^y =\delta \zeta_2^x= -\frac{4 a \gamma^3}{Q + 8 a \gamma^2}\delta g_{xy},\quad\quad\quad
        \delta h^1_{txy}= -\frac{\gamma Q }{\sigma}\delta u^y - \gamma\frac{Q + 4 a \gamma^2}{Q + 8 a \gamma^2} \pa_x\delta g_{xy},
    \end{aligned}
\end{equation}
while for the longitudinal fields it gives
\begin{equation}
    \begin{aligned}
        \label{eq:longitudinalintegrateout}
        \delta \zeta_1^x =&\, -\frac{s_\rho}{2} \delta T  + \frac{4 a \gamma^3 }{Q + 8 a \gamma^2}\delta g_- - \frac{\gamma}{2} \delta g_+ - Qs_\rho \frac{\delta g_+ + 2 \delta g_{tt}}{4 \frac{\partial Q}{\partial T}} , \\
        \delta \zeta_2^y =&\, -\frac{s_\rho}{2} \delta T  - \frac{4 a \gamma^3 }{Q + 8 a \gamma^2}\delta g_- - \frac{\gamma}{2} \delta g_+ -Qs_\rho \frac{\delta g_+ + 2 \delta g_{tt}}{4 \frac{\partial Q}{\partial T}} , \\
        \delta h^2_{txy}=&\, \frac{\gamma Q}{\sigma}\delta u^x + \left( \frac{s_\rho}{2} + \frac{\gamma}{T} \right) \pa_x \delta T +  \frac{Qs_\rho}{4\frac{\partial Q}{\partial T}}\pa_x  \left( \delta g_+ + 2 \delta g_{tt}\right) + \frac{\gamma Q }{2 \left( Q + 8 a \gamma^2 \right)}\pa_x \delta g_+\\
        &\,- \gamma \frac{Q + 4 a \gamma^2}{Q + 8 a \gamma^2} \pa_x \delta g_{yy}  .
    \end{aligned}
\end{equation}
As the operators $\mathcal{O}_I$ are not conserved densities or their associated fluxes in the hydrodynamic theory we are looking to derive, we have set perturbations of their sources to zero in these expressions and from now on.

\paragraph{}In the scalar theory, momentum is no longer conserved due to the sources $\mathcal{J}_I$ that break translational symmetry. Therefore we next integrate out the hydrodynamic velocity fields $\delta u^i$ from the effective theory. This is done by examining the viscoelastic conservation equations for the momentum \eqref{eq:3.52} and \eqref{eq:3.54} which after relabelling are
   \begin{equation} \label{514} \begin{aligned} 
		\left(1+\frac{\chi\sigma}{\gamma^2Q^2}\partial_t\right)\delta u^y  & = - \frac{\chi \sigma}{\gamma^2 Q^2} \pa_t  \delta g_{ty} , \\
        \left(1+\frac{\chi\sigma}{\gamma^2Q^2}\partial_t\right)\delta u^x  & = - \frac{\sigma \chi}{\gamma^2 Q^2}\left(\pa_x \frac{\delta T}{T}+\partial_t\delta g_{tx}-\frac{1}{2} \pa_x \delta g_{tt}\right)  .
    \end{aligned} \end{equation}
To the order in the derivative expansion to which we are working, these can be solved to give the following expressions for the hydrodynamic velocity fields
\begin{equation} \begin{aligned} 
\label{eq:velocitysols}
		\delta u^y =-\frac{\chi\sigma}{\gamma^2Q^2}\pa_t\delta g_{ty}, \quad\quad\quad\quad\quad\quad
        \delta u^x  = - \frac{\sigma \chi}{\gamma^2 Q^2}\left(\pa_x \frac{\delta T}{T}+\partial_t\delta g_{tx}-\frac{1}{2} \pa_x \delta g_{tt}\right).
    \end{aligned} \end{equation}
Note that the fact that there is not a locally conserved momentum or higher-form charge in the scalar formulation of the theory is reflected in the fact that the expressions \eqref{eq:transverseintegrateout}, \eqref{eq:longitudinalintegrateout} and \eqref{eq:velocitysols} for $\delta\zeta_I^i$ and $\delta u^i$ are derivative expansions. This means that there are no non-local terms introduced upon integrating them out.

\paragraph{}To see this explicitly, we look at the the resulting constitutive relations for the energy-momentum tensor. Only one hydrodynamic field -- the local temperature perturbation $\delta T$ -- remains and the only components of the stress tensor that depend on it are $\delta T^t_{\;\; t}$, $\delta T^t_{\;\; x}$, $\delta T_{+}$. The constitutive relations are
   \begin{equation} \begin{aligned} 
   \label{eq:intermediateTscalar}
        \langle\delta T^{t}_{\;\; t}\rangle&=-c_\rho\delta T+ \left(Q\gam^2+\frac{TQ\gam}{2}s_\rho\right)\delta g_+ , \\
         \langle\delta T^{t}_{\;\; x}\rangle & = - \frac{\sigma \chi^2}{\gamma^2 Q^2}\left(\frac{\partial_x\delta T}{T}+\partial_t\delta g_{tx}-\frac{1}{2}\partial_x\delta g_{tt}\right)+\chi\delta g_{tx}, \\
		\langle\delta T_+\rangle & = \left(s-\frac{Q\gam}{2}s_\rho\right) \delta T + \left( \frac{\gamma^4Q \frac{\partial Q}{\partial\gamma^2}}{Q + 2 \gamma^2 \frac{\partial Q}{\partial\gamma^2}} - \zeta \pa_t \right) \delta g_+.
    \end{aligned} \end{equation}
Finally, we use \eqref{eq:EMtensormap} to obtain the constitutive relations of the energy-momentum tensor of the scalar theory. This simply gives $\mathcal{M}$-dependent corrections to the source terms in \eqref{eq:intermediateTscalar}. Specifically
\begin{equation}
    \begin{aligned}
    \label{eq:perturbedTscalar}
        \langle\delta T^{tt}_{\Phi}\rangle&=c_\rho\delta T+\left(-p+sT+2Q\gamma^2-m^2\mathcal{M}\right)\delta g_{tt}+\left(-\frac{TQ\gam}{2}s_\rho-Q\gamma^2+m^2\mathcal{M}\right)\delta g_+, \\
        \langle\delta T^{tx}_{\Phi}\rangle & = - \frac{\sigma \chi^2}{\gamma^2 Q^2}\left(\frac{\partial_x\delta T}{T}+\partial_t\delta g_{tx}-\frac{1}{2}\partial_x\delta g_{tt}\right)+\left(\chi+p-sT-2Q\gamma^2\right)\delta g_{tx}, \\
    \end{aligned}
\end{equation}
and $\delta T_{\Phi +}$ satisfies the tracelessness Ward identity 
\begin{equation}
\begin{aligned}
\label{eq:PerturbedWardscalar}
\langle \delta T^{t}_{\Phi t}\rangle+2\langle \delta T_{\Phi +}\rangle=&\,\left(2s-Q\gam s_\rho-c_\rho\right)\delta T\\
&\,+\left(\frac{2Q\gamma^4\frac{\partial Q}{\partial\gamma^2}}{Q+2\gamma^2\frac{\partial Q}{\partial\gamma^2}}+\frac{TQ\gam}{2}s_\rho+Q\gamma^2-m^2\mathcal{M}-2\zeta\partial_t\right)\delta g_+.
\end{aligned}
\end{equation}
As a consistency check we have verified that the combinations multiplying each $\delta T$ and $\delta g_{\mu\nu}$ term in \eqref{eq:perturbedTscalar} are independent of the double-trace coupling $\mathcal{M}$, which does not exist in the scalar theory. We have further checked that the right hand side of \eqref{eq:PerturbedWardscalar} vanishes when the holographic values of the transport coefficients are substituted in (the first term due to the identity \eqref{eq:shomogeneity}). This is required as the scalar theory is conformal. 

\paragraph{}The remaining dynamical equation is the $t$-component of \eqref{eq:scalarEMconservation}: the energy conservation equation. In the absence of perturbations of the scalar sources, this is
\begin{equation}
\begin{aligned}
\label{eq:scalarenergyconseq}
&\,\partial_t \left(\langle\delta T_\Phi^{tt}\rangle+\left(\chi-m^2\mathcal{M}\right)\delta g_+ +\left(p-\chi-Q\gam^2+m^2\mathcal{M}\right)\delta g_{tt}\right)\\
        &\,+\partial_x \left(\langle\delta T_\Phi^{tx}\rangle+\left(Q\gam^2-p\right)\delta g_{tx}\right)=0.
        \end{aligned}
\end{equation}
The energy conservation equation \eqref{eq:scalarenergyconseq}, combined with the constitutive relations \eqref{eq:perturbedTscalar}, are the hydrodynamics of heat diffusion of \cite{Davison:2014lua}. From the constitutive relations \eqref{eq:perturbedTscalar} we can read off the heat capacity $c$ and thermal conductivity $\kappa$ of the scalar theory in terms of the viscoelastic transport coefficients. The specific heat is defined as the coefficient of $\delta T$ in $\langle\delta T_\Phi^{tt}\rangle$ and the thermal conductivity as minus the coefficient of $\partial_x\delta T$ in $\langle\delta T_\Phi^{tx}\rangle$, and so
\begin{equation}
    c=c_\rho,\quad\quad\quad\quad\quad\quad \kappa=\frac{1}{T}\frac{\sigma\chi^2}{\gamma^2Q^2}.
\end{equation}
The first of these equations is obvious. As it is fixed only by the horizon metric, $s$ is the same function in the scalar and higher-form theories with the identification that $m$ in the former is the charge density in the latter. The heat capacity $c$ in the scalar theory is defined with $m$ fixed (as it is an external source), and so it corresponds to the heat capacity at fixed charge density $c_\rho$ in the latter. Although both $\sigma$ and $\chi$ depend on the double-trace $\mathcal{M}$, they appear in $\kappa$ in a combination that is independent of $\mathcal{M}$.

\paragraph{}This theory has a single hydrodynamic mode: heat diffusion with diffusivity
\begin{equation}
    D_T=\frac{\kappa}{c}=\frac{1}{c_\rho}\frac{\sigma\chi^2}{T\gamma^2Q^2}.
\end{equation}
Substituting in the explicit expressions for the viscoelastic transport coefficients we find agreement with the $c$, $\kappa$ and $D_T$ computed explicitly in \cite{Andrade:2013gsa,Donos:2014cya,Davison:2014lua}. Note that this hydrodynamic mode has a different dispersion relation than the usual diffusion mode \eqref{eq:diffusiondispers} of viscoelastic hydrodynamics. This is because the Hodge map mixes expectation values and sources so that a source-free solution in one formulation is not a source-free solution in the other. However, the diffusive mode of the scalar theory does have a clear origin in the original viscoelastic hydrodynamics. It is the collective excitation satisfying the no-flux condition \eqref{eq:nofluxdispersion}. This is easy to understand: hydrodynamic modes of the scalar theory are solutions with no perturbations of the scalar sources or metric, and from the map \eqref{eq:mattermap} this implies the no-flux condition in the higher-form theory.

\paragraph{}We have now shown how applying the map \eqref{eq:mattermap} and \eqref{eq:EMtensormap} to viscoelastic hydrodynamics yields the simple diffusive hydrodynamics \eqref{eq:perturbedTscalar} and \eqref{eq:scalarenergyconseq} of the scalar theory. Note that going the other way round is not possible: it would require `integrating in' new hydrodynamics degrees of freedom. In other words, to obtain viscoelastic hydrodynamics from the scalar theory would require first augmenting diffusive hydrodynamics to keep careful track of the dynamics of non-conserved operators such as the total momentum, as well as the scalars $\mathcal{O}_I$ and their sources (which under the map \eqref{eq:mattermap} contain the information about the higher-form charges).

\subsection{Momentum-relaxed hydrodynamics}

\paragraph{}For the scalar theory, when $m\ll T$ momentum relaxes slowly and there is an alternative long wavelength, slow-relaxation limit
\begin{equation}
\label{eq:QHlimit}
\partial_\mu\sim m^2/T\ll T,    
\end{equation}
in which a simple effective theory is valid. This is the theory of momentum-relaxed hydrodynamics: at the longest wavelengths $T\partial_\mu \ll m^2$ this reduces to the diffusive hydrodynamics above. However, perturbations of the momentum have a long but finite lifetime $\tau\sim T/m^2$ such that at intermediate wavelengths $T\partial_\mu \gg m^2$, perturbations of the momentum survive and qualitatively alter the dynamics \cite{Davison:2014lua}. We can this obtain momentum-relaxed hydrodynamics at leading order in the long wavelength, slow-relaxation expansion \eqref{eq:QHlimit} via the map from viscoelastic hydrodynamics. To do this, we should retain the momentum conservation equations \eqref{514} within the low energy theory rather than using them to integrate out the velocity perturbations $\delta u^i$. At leading order in the expansion \eqref{eq:QHlimit} these equations become
\begin{equation}
    \left(\tau\partial_t+1\right)\delta u^i=0,
\end{equation}
in the absence of sources. In this equation, the lifetime $\tau$ of perturbations of the state's total momentum is given by the following combination of viscoelastic transport coefficients
\begin{equation}
    \tau=\lim_{m\rightarrow0}\frac{\chi\sigma}{\gamma^2Q^2}=\frac{4\pi T}{m^2}.
\end{equation}
In combination with the small $m$ limit of the energy dynamics above, this yields the leading order momentum-relaxed hydrodynamics \cite{Davison:2014lua}. The general theory of momentum-relaxed hydrodynamics has been extended beyond this leading order \cite{Gouteraux:2023uff} and describes the dynamics of the scalar theory \cite{Davison:2015bea,Blake:2015epa}. However, we are not able to obtain the subleading corrections from applying the Hodge map to the viscoelastic hydrodynamics of Section \ref{sec:HydroSec}. There are first-order terms in the momentum-relaxed hydrodynamics expansion \eqref{eq:QHlimit} that arise from second order terms in the hydrodynamic expansion of viscoelastic hydrodynamics. Since we only worked to first order in viscoelastic hydrodynamics we cannot reproduce these.

\section{Discussion}
\label{sec:discussion}

\paragraph{}We have shown analytically that small amplitude and long wavelength perturbations of the black brane \eqref{eq:BHmetric} charged under a 2-form potential are governed by a simple effective theory: relativistic, viscoelastic hydrodynamics. 

\paragraph{}In this regime, the theory of viscoelastic hydrodynamics contains 5 independent, undetermined functions of the temperature and chemical potential: the pressure, shear modulus, charge conductivity, and shear and bulk viscosities. The pressure, and its associated thermodynamic derivatives (the entropy density, charge density and bulk modulus) were first computed in \cite{Grozdanov:2018ewh,Armas:2019sbe,Armas:2020bmo}. Our results for the other transport coefficients are new. The bulk viscosity vanishes, even though the double-trace coupling explicitly $\mathcal{M}$ breaks conformal symmetry. An explicit expression for the conductivity is given in equation \eqref{eq:transportcoeffsholoresults}. The shear modulus and shear viscosity are given by solutions to the very simple time-independent differential equation \eqref{eq:Theta1Eq} that we have solved analytically in various limits. The gravitational dynamics are completely described by the lower dimensional hydrodynamic theory: for example, our results guarantee that the quasinormal modes of the black brane are simply the viscoelastic hydrodynamic modes described in Section \ref{sec:hydromodes}. The first obvious generalisation of our work is to go beyond the small amplitude approximation, as is done in the original fluid/gravity correspondence \cite{Bhattacharyya:2007vjd,Erdmenger:2008rm,Banerjee:2008th}, and show that the complete long wavelength gravitational dynamics can be described by a lower-dimensional viscoelastic theory. In \cite{Baggioli:2019mck,Baggioli:2020qdg,Pan:2021cux,Baggioli:2023dfj}, aspects of the non-linear shear response of a massive gravity-like model were investigated numerically.

\paragraph{}The double-trace coupling $\mathcal{M}$ was introduced in \cite{Grozdanov:2018ewh}, where it was pointed out that the equilibrium state is unstable unless $\mathcal{M}>r_h$. Our analysis shows that there are numerous perturbative, long wavelength instabilities even above this threshold and identified \eqref{eq:longitudinalstabilitythreshold} as the minimal value of the coupling for stability. This first instability to onset is driven by perturbations of the incoherent density defined in \eqref{eq:incoherentdensitydef}. It would be very interesting to identify the equilibrium state at the endpoint of this instability.

\paragraph{}A second effect of the double-trace coupling is that it produces additional long-lived modes when $\mathcal{M}/T$ is large \cite{Grozdanov:2018ewh}. These could be interpreted as Goldstones that decouple from the thermal bath at low temperatures, since in the scalar formulation of viscoelasticity the coupling is that of the Goldstone kinetic term.\footnote{In a similar way to how the corresponding mode in holographic theories with the higher-form symmetry of magnetohydrodynamics is interpreted as a photon \cite{Hofman:2017vwr,Grozdanov:2017kyl}. See \cite{Pinheiro:2025fqg} for further examples like this.} It would be very interesting to use our results to see whether there is a controlled limit of large $\mathcal{M}$ along with small wavelengths and frequencies for which a quasi-hydrodynamic theory \cite{Grozdanov:2018fic} incorporating these additional modes applies.

\paragraph{}Although we focused on the simplest example of a black brane solution charged under 2-form potentials, the long wavelength perturbations of any solution of this type should be obey the equations of relativistic viscoelastic hydrodynamics. We expect that our analysis can be adapted and generalised to a much wider class of such theories. Although in general we don't anticipate being able to solve the bulk linearised equations of motion in the same detail as we have done here, we expect the techniques of \cite{Donos:2022uea} to be sufficient to prove the equivalence to viscoelastic hydrodynamics and formally extract the transport coefficients. A simple family of extensions is to allow the bulk coupling of the gauge fields to be set by a running scalar field, rather than simply a constant. Firstly, using more general theories of this type it may be possible to construct equilibrium states with better stability properties. Secondly, this will help clarify which aspects of transport in holographic theories are genuinely generic. We anticipate that a running gauge field coupling will generate a non-zero bulk viscosity and when the gauge coupling becomes small in the interior we anticipate that new (generalised) symmetries may emerge (analogously to the zero-form case \cite{Davison:2018ofp,Davison:2018nxm,Davison:2022vqh}) and lead to a much richer effective theory. Other types of generalisations are to incorporate an additional 0-form U(1) symmetry (see \cite{Armas:2020bmo}) or to consider non-isotropic equilibrium states, whose small amplitude perturbations are sensitive to a larger set of the effective theory's transport coefficients due to the reduced symmetry.

\paragraph{}Finally, we have focused on the case of isotropic, viscoelasticity in (2+1)-dimensions as it is arguably the simplest example to study holographically. But the methods used here can easily be adapted to other physically important cases including magnetohydrodynamics \cite{Grozdanov:2016tdf,Hernandez:2017mch,Armas:2018ibg,Armas:2018atq,Armas:2018zbe} and superfluidity \cite{Delacretaz:2019brr}. In the case of magnetohydrodynamics, the recent work \cite{Frangi:2025ykb} highlighted the importance of incorporating perturbations of the bulk metric as we have done. We expect that our analysis can be adapted to incorporate states with only approximate higher-form symmetries: the corresponding hydrodynamic theories have numerous phenomenological applications and have been investigated in \cite{Delacretaz:2021qqu,Armas:2021vku,Baggioli:2022pyb,Armas:2022vpf,Armas:2023tyx}. Some holographic steps in this direction were taken in \cite{Pinheiro:2025fqg} (see also \cite{Baggioli:2014roa,Alberte:2015isw,Alberte:2017cch,Jokela:2017ltu,Andrade:2017ghg,Andrade:2018gqk,Amoretti:2018tzw,Donos:2019tmo,Romero-Bermudez:2019lzz,Donos:2019hpp,Amoretti:2019kuf,Baggioli:2020nay,Andrade:2020hpu,Andrade:2022udb} for related previous studies).

\acknowledgments

We are very thankful to Sa\v{s}o Grozdanov and Nick Poovuttikul for sharing the numerical results of \cite{Grozdanov:2018ewh} with us, which we have reproduced in Figure \ref{fig:TransverseModes1}. We are also grateful to Jay Armas, Matteo Baggioli and especially Akash Jain for very helpful discussions. The work of RD was supported in part by
the STFC Ernest Rutherford Grant ST/R004455/1. AOP was supported by an EPSRC Studentship.

\appendix

\section{Holographic dictionary}
\label{app:holodictionary}

\paragraph{}In this Appendix we summarise the holographic dictionaries that we use to relate expectation values and sources of operators in the dual QFT to the near-boundary expansion of solutions to the bulk equations of motion. Throughout, we work in radial gauge near the boundary. The expressions in this Appendix are are consistent with those in \cite{Andrade:2013gsa,Grozdanov:2018ewh,Armas:2019sbe}.

\subsection{Higher-form theory}

\paragraph{}First we consider the theory with a bulk three-form field strength and action \eqref{lag}. Neglecting horizon contributions, the on-shell variation of this action is the boundary term
\be \begin{aligned} 
	\del S =\frac{1}{2} \int d^3x\sqrt{-g} \left(\braket{T^{\mu \nu}} \del g_{\mu \nu} - \braket{J_I^{\mu \nu}} \del b^I_{\mu \nu}\right),
\end{aligned} \ee
after taking the cutoff $\Lambda\rightarrow\infty$, where
\begin{equation}
\begin{aligned}
\label{eq:generalholodicHF}
\braket{T^{\mu \nu}} & = \lim_{\Lam \to \infty} \frac{2 \Lam^5}{\sgam} \left( \frac{\pa \lag_{bulk}}{\pa (\pa_r G_{\mu \nu}) } + \frac{\pa \lag_{ct}}{\pa \gam_{\mu \nu}} \right) \bigg\vert_{r = \Lam}, \quad g_{\mu \nu} = \lim_{\Lam \to \infty} \Lam^{-2} \gam_{\mu \nu} \big\vert_{r = \Lam},   \\  
\braket{J_I^{\mu \nu}} & = \lim_{\Lam \to \infty} \Lam^3 \gam^{\mu \rho} \gam^{\nu \sig} \Pi^I_{\rho \sig} \big\vert_{r = \Lam},\quad\quad\quad\quad\quad\quad\quad b^I_{\mu \nu} = \lim_{\Lam \to \infty} \left( B^I_{\mu \nu} - \frac{\Pi^I_{\mu \nu}}{\ka (\Lam)} \right) \Bigg\vert_{r = \Lam} .
\end{aligned}    
\end{equation}
Here, $\lag_{ct}$ is the integrand of the counterterm action \eqref{eq:counterterm} and $\lag_{bulk}$ arises from integrating the bulk action by parts to cancel the GHY term:
\begin{equation}
    \lag_{bulk} =\sqrt{-G}\left( \left( K_{ab} G^{ab} \right)^2 - K_{ab} K_{cd} G^{ac} G^{bd} + \mathcal{R} [\gam] + 6 - \frac{G^{ad} G^{be} G^{cf}}{12} H^I_{abc} H^I_{def}\right),
\end{equation}
where $K_{ab}=\mathcal{L}_n\gamma_{ab}/2$ is the extrinsic curvature. Explicitly, 
\begin{equation}
    	\frac{\pa \lag_{bulk}}{\pa (\pa_r G_{\mu \nu}) } = 2 \sgam \gam^{\mu [\nu} \gam^{a] b} K_{ab}.
\end{equation}

\paragraph{}Evaluating the expressions \eqref{eq:generalholodicHF} on the equilibrium black hole solution \eqref{eq:BHmetric} gives the results for the equilibrium expectation values quoted in the main text. For the small amplitude perturbations \eqref{2.8} around this equilibrium solution, the expressions \eqref{eq:generalholodicHF} become
\begin{equation}
\begin{aligned} 
	\del g_{\mu \nu} = \lim_{\Lam \to \infty} \Lam^{-2} \hat{g}_{\mu \nu} \big\vert_{r = \Lam},\quad\quad\quad\quad\quad\quad
	\del b^I_{\mu \nu} = \lim_{\Lam \to \infty} \left( \hat{b}^I_{\mu \nu} + (\mathcal{M} - \Lam) \hat{h}^I_{r \mu \nu} \right) \big\vert_{r = \Lam},
\end{aligned}
\end{equation}
for the sources,
\begin{equation}
    \begin{aligned}
\braket{\delta J_I^{\mu \nu}} 
        & = \lim_{\Lam \to \infty} \left( r^2 \hat{h}_I^{r \mu \nu} + \nn r^4 \ep_{IJ} \hat{\ep}^{r \mu \nu \al} \del^J_\beta \hat{g}^\beta_\al - \braket{\bar{J}_I^{\mu \nu}} \bar{g}^{\rho \sig} \hat{g}_{\rho \sig}  \right) \bigg\vert_{r = \Lam},        
    \end{aligned}
\end{equation}
for the higher-form currents, and finally
\begin{equation}
    \begin{aligned}
    \braket{\del T^t {}_t} & = \lim_{\Lam\rightarrow\infty} \Lam^3 \left.\left( \frac{2 \pa_r \hat{g}_+}{r \sqrt{f}} - \nn \frac{r^2 \hat{h}_1^{rtx} + r^2 \hat{h}_2^{rty} + \nn \hat{g}^t_t + \nn \hat{g}_+}{\ka (\Lam) r^2}\right)\right|_{r=\Lam}, \\
    \braket{\del T^t {}_x} & = \lim_{\Lam\rightarrow\infty} \Lam^3 \left.\left( \frac{\pa_r \hat{g}^x_t}{r f^{3/2}} + \frac{\nn \hat{h}_2^{rxy}}{\ka (\Lam) f}\right)\right|_{r=\Lam}, \\
    \braket{\del T^t {}_y} & = \lim_{\Lam\rightarrow\infty} \Lam^3 \left.\left( \frac{\pa_r \hat{g}^y_t}{r f^{3/2}} - \frac{\nn \hat{h}_1^{rxy}}{\ka (\Lam) f}\right)\right|_{r=\Lam}, \\
    \braket{\del T^x {}_y} & = \lim_{\Lam\rightarrow\infty} \Lam^3 \left.\left( - \frac{\pa_r \hat{g}^x_y}{r \sqrt{f}} - \nn \frac{r^2 \hat{h}_2^{rtx} + r^2 \hat{h}_2^{rty} + \nn \hat{g}^x_y}{\ka (\Lam) r^2}\right)\right|_{r=\Lam}, \\
    \braket{\del T_+} & = \lim_{\Lam\rightarrow\infty} \Lam^3 \left.\left( \frac{\pa_r \hat{g}_+ + \pa_r \hat{g}^t_t}{r \sqrt{f}} \right)\right|_{r=\Lam}, \\
    \braket{\del T_-} & = \lim_{\Lam\rightarrow\infty} \Lam^3 \left.\left( \frac{- \pa_r \hat{g}_-}{r \sqrt{f}} + \nn \frac{r^2 \hat{h}_1^{rtx} - r^2 \hat{h}_2^{rty} - \nn \hat{g}_-}{\ka (\Lam) r^2}\right)\right|_{r=\Lam},    
    \end{aligned}
\end{equation}
for the energy-momentum tensor, where $\bar{g}_{\mu\nu}$ and $\langle \bar{J}^{\mu\nu}_I\rangle$ on the right hand side denote the equilibrium values for these quantities.

\subsection{Scalar theory}

\paragraph{}Now we turn to the scalar theory with action \eqref{eq:Einsteinscalarbulkaction} and Dirichlet boundary conditions. This boundary condition requires the counterterm
\begin{equation}
    S_{\Phi,ct}=\int d^3x\sgam \left( \frac{F^\mu_I F_\mu^I}{2} - 4 \right).
\end{equation}
in addition to the usual Gibbons-Hawking-York term. Notice that there is no analogue of the double-trace coupling $\mathcal{M}$ in this theory. This means that we use the usual Dirichlet boundary condition for the scalar field. More specifically, the on-shell variation of the renormalised action (after removing the cutoff $\Lambda\rightarrow\infty$) is
\begin{equation} \begin{aligned} 
\del S_\Phi & = \int d^3x \sqrt{- g} \left( \frac{\braket{T^{\mu \nu}_\Phi} \del g_{\mu \nu}}{2} - \braket{\mathcal{O}^I} \del \mathcal{J}_I \right),
\end{aligned} \end{equation}
where
\begin{equation}
\begin{aligned} 
\label{eq:scalarHolodic}
	\braket{T_{\Phi}^{\mu \nu}} & = \lim_{\Lam \to \infty} \frac{2 \Lam^5}{\sgam} \left( \frac{\pa \lag_{\Phi,bulk}}{\pa (\pa_r G_{\mu \nu}) } + \frac{\pa \lag_{\Phi,ct}}{\pa \gam_{\mu \nu}} \right) \bigg\vert_{r = \Lam}, \\
	\braket{\mathcal{O}^I} & = \lim_{\Lam \to \infty} \Lam^3 \left( n^a F_a^I + \na_\mu F^\mu_I \right)  \bigg\vert_{r = \Lam}, \\
	\mathcal{J}_I & = \lim_{\Lam \to \infty} \Phi_I \big\vert_{r = \Lam}, 
\end{aligned}
\end{equation}
and the boundary metric $g_{\mu\nu}$ given by the same expression \eqref{eq:generalholodicHF} as in the higher-form theory. The covariant derivative on the right hand side of equation \eqref{eq:scalarHolodic} is that compatible with $G_{ab}$. $\lag_{\Phi,ct}$ is the integrand of $S_{\Phi,ct}$ and 
\begin{equation}
\lag_{bulk}^\Phi = \sqrt{-G} \left( \left( K_{ab} G^{ab} \right)^2 - K_{ab} K_{cd} G^{ac} G^{bd} + \mathcal{R} [\gam] + 6 - \frac{1}{2} F^I_a F_I^a \right).    
\end{equation}

\paragraph{}Applying the map \eqref{eq:hodgetransform} to the right hand side of the higher-form dictionary \eqref{eq:generalholodicHF} and then comparing the results to the scalar dictionary \eqref{eq:scalarHolodic} gives the relations \eqref{eq:mattermap} and \eqref{eq:EMtensormap} between the sources and expectation values in the two field theories used in the main text.

\bibliographystyle{JHEP}
\bibliography{ViscoHoloBib}
\end{document}